\documentclass[12pt,english]{article}
\usepackage{mathptmx}

\usepackage[T1]{fontenc}
\usepackage[latin9]{inputenc}
\usepackage{geometry}
\usepackage{color}
\usepackage{amsbsy}
\usepackage{amstext}
\usepackage{amssymb}
\usepackage{graphicx}
\usepackage{esint}

\makeatletter

\providecommand{\tabularnewline}{\\}

\newcommand{\lyxaddress}[1]{
\par {\raggedright #1
\vspace{1.4em}
\noindent\par}
}

\makeatother

\usepackage{babel}
\begin{document}

\title{Control of resistive wall modes in a cylindrical tokamak with plasma
rotation and complex gain}

\author{D.~P.~Brennan$^{*}$ and J.~M.~Finn$^{\dagger}$}

\maketitle

\lyxaddress{$^{*}$Department of Astrophysical Sciences, Princeton University,
Princeton, NJ 08544; $^{\dagger}$ Applied Mathematics and Plasma
Physics, Theoretical Division, Los Alamos National Laboratory, Los
Alamos, NM 87545}
\begin{abstract}
Feedback stabilization of magnetohydrodynamic (MHD) modes is studied
in a cylindrical model for a tokamak with resistivity, viscosity and
toroidal rotation. The control is based on a linear combination of
the normal and tangential components of the magnetic field just inside
the resistive wall. The feedback includes complex gain, for both the
normal and for the tangential components, and the imaginary part of
the feedback for the former is equivalent to plasma rotation. The
work includes (1) analysis with a reduced resistive MHD model for
a tokamak with finite $\beta$ and with stepfunction current density
and pressure profiles, and (2) computations with full compressible
visco-resistive MHD and smooth decreasing profiles of current density
and pressure. The equilibria are stable for $\beta=0$ and the marginal
stability values $\beta_{rp,rw}<\beta_{rp,iw}<\beta_{ip,rw}<\beta_{ip,iw}$
(resistive plasma, resistive wall; resistive plasma, ideal wall; ideal
plasma, resistive wall; ideal plasma, ideal wall) are computed for
both cases. The main results are: (a) imaginary gain with normal sensors
or plasma rotation stabilizes below $\beta_{rp,iw}$ because rotation
supresses the diffusion of flux from the plasma out through the wall
and, more surprisingly, (b) rotation or imaginary gain with normal
sensors destabilizes above $\beta_{rp,iw}$ because it prevents the
feedback flux from entering the plasma through the resistive wall
to form a virtual wall. The effect of imaginary gain with tangential
sensors is more complicated but essentially destabilizes above and
below $\beta_{rp,iw}$. A method of using complex gain to optimize
in the presence of rotation in the $\beta>\beta_{rp,iw}$  regime
is presented.
\end{abstract}

\section{Introduction}

Feedback stabilization of magnetohydrodynamic (MHD) modes with plasma
resistivity and a resistive wall in tokamaks has received recent attention
particularly because of the need to control disruptions\cite{bondeson-ward,betti-freidberg,finn_rw1,finn_rw2,finn_rw3,finn_rw4,Bhattacharyya,Boozer-slow-rotation,GA-res-wallI,fitzpatrick-aydemir,garofalo_2002,Liu2006}.
Studies have also been performed for reversed field pinches (RFPs)\cite{Bishop,RichardsonFinnDelzanno,Sassenberg}.
Earlier studies in tokamak geometry\cite{Finn2004,bondeson_pop,Tangential-normalChuGlasser,bondeson2001,Chu2004,mode-control,Pustovitov2002}
investigated sensing either the radial or the poloidal component of
the magnetic field, concluding that it is better to sense the poloidal
component, and that the latter measurement is of more use inside the
wall\cite{Pustovitov2002,Finn2004}. Results in Refs.~\cite{Tangential-normalChuGlasser,Finn2004}
suggested that the advantages of tangential sensing are due to the
fact that it is less sensitive to sensors that detect sidebands or
feedback coils that excite sidebands.

In Ref.~\cite{Finn2006} studies were performed with of \emph{both}
the radial and poloidal components (radial and \emph{toroidal} components
in the RFP context) but with idealized (single Fourier component)
coils. The results showed that this approach has useful advantages
over control based on either sensor alone. Whereas feedback based
on sensing either field component alone is limited to the marginal
stability point for resistive plasma modes with an ideal wall, feedback
based on sensing both components can stabilize up to the ideal plasma
- ideal wall limit. These results were presented in Ref.~\cite{Finn2006},
which used a very simple qualitative model based on reduced resistive
MHD\cite{Strauss} for the plasma dynamics. More recent investigations
in full visco-resistive MHD in a cylindrical model for RFPs\cite{RichardsonFinnDelzanno,Sassenberg}
have shown this ability to stabilize close to the ideal plasma - ideal
wall limit, depending on the plasma viscosity and resistivity. In
Ref.~\cite{Sassenberg} the work in Ref. \cite{RichardsonFinnDelzanno}
was extended to a model measuring the radial component and \emph{two}
tangential components of the magnetic field, again in RFP geometry.
This work included the presence of two walls, the (inner) vacuum vessel
and a better conducting external copper shell, with sensors between
the two walls, as suggested by the configuration of the RFX-mod facility\cite{RFX-mod}.
The results of this study also showed the possibility of stabilizing
close to the ideal plasma - ideal wall limit, depending on the plasma
viscosity and the placement of the sensors, and that the second tangential
component (toroidal in tokamak geometry and poloidal in RFP geometry)
is not important. The RFX-mod facility has the capability of sensing
both the normal and toroidal components and applying a \emph{pre-specified}
linear combination of these\cite{RFXPiron,RFXAuthors}. In the theoretical
work in Refs.~\cite{RichardsonFinnDelzanno,Sassenberg}, the normal
and tangential components were considered independent. The RFP results
with $\beta=0$ were parameterized in terms of the critical values
of the equilibrium current density at the magnetic axis, i.e.~$\lambda_{0}=(j_{||}/B)(r=0)$,
namely $\lambda_{rp,rw}<\lambda_{rp,iw}<\lambda_{ip,rw}<\lambda_{ip,iw}$.
These four values of $\lambda_{0}$ are, respectively the current
limits for resistive plasma, resistive wall; resistive plasma, ideal
wall; ideal plasma, resistive wall; and ideal plasma, ideal wall.
The inner inequality $\lambda_{rp,iw}<\lambda_{ip,rw}$ was observed
to hold\cite{RichardsonFinnDelzanno,Sassenberg} for all RFP equilibria
investigated. The other inequalities must always hold.

In this paper we investigate linear stability in a finite-$\beta$
cylindrical model with tokamak-like profiles, namely large toroidal
aspect ratio $R/a$, large toroidal field $B_{z}\sim(R/a)B_{\theta}$
and decreasing profiles of current density $j_{z0}(r)$ and pressure
$p_{0}(r)$. The decreasing $j_{z0}(r)$ profile leads to a monotonically
increasing profile of the safety factor $q(r)=rB_{z0}/RB_{\theta0}(r)$
with $q\sim1$. We consider equilibria which are stable for zero pressure
and characterize the stability properties without feedback in terms
of the four marginal values of $\beta_{0}=2p_{0}(0)/B_{z0}(0)^{2}$,
namely $\beta_{rp,rw}<\beta_{rp,iw}<\beta_{ip,rw}<\beta_{ip,iw}$,
analogous to the values of $\lambda=j_{||}/B$ at $r=0$ in the RFP
studies. (As in the RFP studies, the middle inequality, which does
not hold in general, has been observed to hold for all the equilibria
we considered.) We again investigate the behavior with feedback proportional
to the radial and poloidal magnetic field components, with gain factors
$G$ and $K$, respectively. We also include toroidal plasma rotation
and complex gain\cite{bondeson_prl} for both the normal component
and the tangential component, i.e.~$G$ and $K$. (Complex gain is
attained by shifting the phase of the actuator coils relative to the
sensor coils.) In Ref.~\cite{FinnChacon1} it was argued that, in
cylindrical geometry with a single $k_{z}$, the imaginary part $G_{i}=\text{Im}G$
is equivalent to rotation of the wall, which is in turn equivalent
to rigid rotation of the plasma. 

An aspect of our studies worth emphasizing is the inclusion of plasma
resistivity as well as wall resistivity. This inclusion introduces
two important marginal stability parameters, namely $\beta_{rp,rw}$
and $\beta_{rp,iw}$, that are absent in ideal MHD. Also, above the
latter limit, modes are unstable but grow on the wall time $\tau_{w}$
and are therefore sensitive to plasma resistivity and react differently
to plasma rotation.

As in the RFP control studies, the control is applied at a surface
\emph{external} to the resistive wall. This is in spite of the fact
that in some current devices actuators are located inside the wall.
Our focus on control applied outside the wall is motivated by the
obvious potential problems of internal control coils, as well as the
results shown here, indicating the possibility of stabilizing well
above the resistive plasma-ideal wall threshold.

In Sec.~2 we describe the cylindrical MHD equilibria used in the
analytic and the numerical studies. In the former case, the simplified
equilibrium has large $B_{z0}$ and stepfunction models for $j_{z0}(r)$
and $p_{0}(r)$. In the latter the equilibrium is specified by smooth
functions for $j_{z0}(r)$ and $p_{0}(r)$.

In Sec.~3 we describe the methods used to analyze the stability of
the simplified model, as well as the full MHD model used to study
the stability of the smooth profile equilibria. In the former we use
reduced resistive MHD\cite{Strauss} in the viscoresistive (VR) regime,
with a single resistive wall and control applied at a wall external
to the resistive wall. We also formulate the problem with a layer
in the resistive-inertial (RI) regime for comparison. The use of reduced
MHD with plasma resistivity and stepfunction profiles enables us to
obtain analytic results for which the various physical effects in
the presence of plasma rotation and feedback with complex gains are
transparent. The studies in full MHD enable us to determine how well
the results of the simplified model represent those of the full model. 

In Sec.~4 we show results using both models. We first present studies
of the stability properties, in particular the four values $\beta_{rp,rw},\,\beta_{rp,iw},\,\beta_{ip,rw},\,\beta_{ip,iw}$,
without rotation or gain. We then present results with real gains
$G=G_{r}$ and $K=K_{r}$, with increasing $\beta_{0}\equiv2p_{0}(0)/B_{z0}(0)^{2}$.

In Sec.~5 we show results including rotation $\Omega$ and complex
gain $G_{i}$, with $K_{i}=0$. The main result is that the behavior
depends on the value of $\beta_{0}$ relative to $\beta_{rp,iw}$,
the resistive plasma - ideal wall threshold. For $\beta_{0}<\beta_{rp,iw}$
plasma rotation $\Omega$ and $G_{i}$ (equivalent to wall rotation
$\Omega_{w}$ and therefore equivalent to plasma rotation in the opposite
direction) are stabilizing, leading to a larger region of stability
in the $(K,G)$ space. This is because rotation of the plasma relative
to the wall suppresses the resistive wall mode by preventing the flux
from diffusing through the wall. For $\beta_{0}>\beta_{rp,iw}$, rotation
relative to the wall is found to be \emph{destabilizing}: in this
regime, the resistive plasma mode is unstable even with an ideal wall,
and for the feedback to succeed the flux needs to diffuse through
the wall in order to form a virtual wall\cite{Bishop} inside the
actual wall\cite{Bishop,Finn2006,RichardsonFinnDelzanno,Sassenberg}.
The stabilizing effect of rotation or $G_{i}$ for $\beta_{0}<\beta_{rp,iw}$
and the destabilizing effect for $\beta_{0}>\beta_{rp,iw}$ is similar
to the dependence on the wall time observed in Ref.~\cite{Finn2006}.
For finite plasma rotation $\Omega\neq0$ the optimum value of $G_{i}$
when $\beta_{0}>\beta_{rp,iw}$ is that value which makes the equivalent
wall rotation $\Omega_{w}$ equal to the plasma rotation $\Omega$,
allowing the fastest penetration of the flux from the feedback coils. 

In Sec.~6 we study the effects of $K_{i}$. It is also found that
in this regime there is no simple equivalence between $K_{i}$ and
plasma rotation, although $K_{i}$ affects the modes in a manner which
has some similarity to rotation. Increasing $|K_{i}|$ is destabilizing
for both $\beta_{0}<\beta_{rp,iw}$ and for $\beta_{0}>\beta_{rp,iw}$,
so it is not possible to interpret $K_{i}$ in terms of equivalent
wall rotation. There is also an optimal value of $K_{i}$ for $\Omega\neq0,\, G_{i}=0$,
both above and below $\beta_{rp,iw}$. For $\beta_{0}>\beta_{rp,iw}$
this behavior is similar to that for $G_{i}$ for reduced MHD, but
is more complicated for full MHD. For $\beta_{0}<\beta_{rp,iw}$ rotation
is stabilizing and the optimal value of $K_{i}$ can generally expand
the stable region.

The change in behavior across $\beta_{0}=\beta_{rp,iw}$ for all values
of $\Omega,\,\, G_{i}$ and $K_{i}$ indicates the importance of plasma
modeling including plasma resistivity.

In Sec.~7 we summarize and discuss the results presented, particularly
the possibility of stabilization well above $\beta_{rp,iw}$ by optimization
using complex gain. We also emphasize the fact that the simple analytic
modeling predicts qualitatively most of the phenomena found by the
more complete full MHD treatment, and that resistive MHD modeling
is necessary to obtain these conclusions because the modes are resonant.

\section{Equilibria}

The equilibrium for the simplified reduced MHD model is specified
in terms of decreasing stepfunction profiles of current density and
pressure, i.e.

\begin{equation}
\begin{array}{c}
B_{\theta0}(r)=r\,\,\,\,\text{for}\,\,\,\, r<a_{1}\\
=\frac{a_{1}^{2}}{r}\,\,\,\,\text{for}\,\,\,\, r>a_{1}\\
j_{z0}(r)=2\Theta(a_{1}-r)\\
B_{z0}(r)=B_{0}\,\,\,\,=\,\text{const}\\
p_{0}(r)=p_{0}(0)\Theta(a_{2}-r).
\end{array}\label{eq:Equilibrium}
\end{equation}
Length scales are relative to $r_{w}$ and time scales to $r_{w}/v_{A}$,
with $v_{A}$ based on the nominal equilibrium poloidal field $B_{\theta0}'(0)r_{w}$,
so $B_{\theta}$ is normalized to have $B_{\theta0}'(0)=r_{w}=1$.
The major radius $R$ satisfies $\epsilon\equiv r_{w}/R\ll1$. For
equilbria in reduced MHD, we take $B_{\theta}\sim\epsilon B_{z}$
and $p\sim B_{\theta}^{2}\sim\epsilon^{2}B_{z}^{2}$. It follows that
$B_{z}B_{z}'\sim\epsilon^{2}B_{z}^{2}$, so that at the steps at $r=a_{1}$
and $r=a_{2}$ we have $\Delta B_{z0}\sim\epsilon^{2}B_{z0}$. This
means that it is consistent to treat $B_{z}$ as uniform and still
have force balance in equilibrium. The $q$ profile is given by
\[
q(r)=q(0)\,\,\,\,\text{for}\,\,\,\, r<a_{1},\,\,\,\,\,\,=q(0)\frac{r^{2}}{a_{1}^{2}}\,\,\,\,\text{for}\,\,\,\, r>a_{1},
\]
where $q(0)=B_{0}/R$ and $R$ is the major radius. The modes behave
as $e^{im\theta+ikz}$ with $k=-n/R$ and $n=1$. We assume $q(0)<m/n$
but $q(a_{2})>m/n$, so that the four radii $a_{1},\, r_{t},\, a_{2},r_{w}$
satisfy $a_{1}<r_{t}<a_{2}<r_{w}$. Here, $r_{t}$ is the radius of
the mode rational surface (tearing layer), which satisfies $q(r_{t})=m/n$,
and $r_{w}$ is the radius of the resistive wall. See Fig.~1a. We
also have a control surface at $r=r_{c}>r_{w}$. Plasma rotation is
represented by a uniform equilibrium toroidal velocity $u_{z0}$.

The equilibrium used for the numerical studies in full MHD is specified
by the toroidal current density $j_{z0}(r)$ and the pressure $p_{0}(r).$
The current density used is the `flattened model' of Ref.~\cite{FRS},
with pressure $p_{0}(r)$ added having a profile similar to $j_{z0}(r)$.
Specifically, we take
\[
B_{\theta0}(r)=\frac{r}{\left(1+(r/a_{1})^{2\nu}\right)^{1/\nu}}
\]
with $\nu=4$, again normalized to have $B_{\theta0}'(0)=1$. Hence
we have
\begin{equation}
j_{z0}(r)=\frac{2}{\left(1+(r/a_{1})^{2\nu}\right)^{(\nu+1)/\nu}}.\label{eq:jz0-specification}
\end{equation}
For the pressure we take a similar form with $\nu=6$, 
\begin{equation}
p_{0}(r)=\frac{p_{0}(0)}{\left(1+(r/a_{2})^{2\nu}\right)^{(\nu+1)/\nu}}.\label{eq:p0-specification}
\end{equation}
Radial force balance $j_{\theta0}B_{z0}-j_{z0}B_{\theta0}=p_{0}'(r)$
gives the toroidal field by
\[
\frac{B_{z0}^{2}}{2}=\frac{B_{0}^{2}}{2}+p_{00}-p_{0}(r)-\int_{0}^{r}j_{z0}(r')B_{\theta0}(r')dr'.
\]
We use the integration constant $B_{0}=B_{z0}(0)$ to specify $q(0)$,
where $q(r)=rB_{z0}(r)/RB_{\theta0}(r)$, i.e.~$q(0)=B_{0}/R$. Here,
as above, the toroidal aspect ratio is $R/r_{w}$. These equilibrium
quantities are shown in Fig.~1b. The equilibrium velocity $u_{z0}$
is again taken to be uniform.

\section{Linear models}

In this section we describe the linear models used to compute the
stability of the stepfunction and smooth equilibria introduced in
the last section. In the first case, we do asymptotic matching with
a viscoresistive (VR) or resistive inertial (RI) inner layer model,
with outer regions derived from finite $\beta$ reduced ideal MHD
without inertia. In the second case, we solve the complete resistive
MHD equations with viscosity, compressional effects and parallel dynamics.

\subsection{Simplified linearized MHD model}

The simplified model for treating resistive MHD modes in a large aspect
ratio cylinder model for a tokamak with a resistive wall uses reduced
MHD\cite{Strauss} with plasma resistivity, and with stepfunction
profiles as described in Sec.~2. In this model the linear dynamics
is described entirely in terms of the perturbed flux function $\tilde{\psi}=\tilde{A}_{z}$,
with $\mathbf{\tilde{B}}=\nabla\tilde{\psi}(r,\theta,z)\times\hat{\mathbf{e}}_{z}$
-- the toroidal field is not perturbed. The (perpendicular) velocity
is given in terms of the perturbed streamfunction by $\tilde{\mathbf{v}}_{\perp}=\nabla\tilde{\phi}\times\hat{\mathbf{e}}_{z}$.
The reduced MHD equations are given in the outer region (ideal MHD,
zero inertia) by
\begin{equation}
0=iF(r)\nabla_{\perp}^{2}\tilde{\psi}-\frac{im}{r}j_{0}'(r)\tilde{\psi}+\frac{2imB_{\theta0}^{2}(r)}{B_{0}^{2}r^{2}}\tilde{p},\label{eq:reduced-1}
\end{equation}
\begin{equation}
\gamma_{d}\tilde{\psi}=iF(r)\tilde{\phi}\,\,\,\,\,\,\,\,\,\,\,\,\,\,\,\,\gamma_{d}\tilde{p}=-\frac{im}{r}p_{0}'(r)\tilde{\phi},\label{eq:reduced-2}
\end{equation}
where $F(r)=mB_{\theta0}(r)+kB_{0}$ and $\gamma_{d}$ is the Doppler
shifted growth rate $\gamma+iku_{z0}=\gamma+i\Omega$. We obtain
\begin{equation}
\nabla_{\perp}^{2}\tilde{\psi}=\frac{mj_{z0}'(r)}{rF(r)}\tilde{\psi}+\frac{2m^{2}B_{\theta0}^{2}(r)p_{0}'(r)}{B_{0}^{2}r^{3}F(r)^{2}}\tilde{\psi}\label{eq:OuterRegionEq}
\end{equation}
\begin{equation}
=-A\delta(r-a_{1})\tilde{\psi}-B\delta(r-a_{2})\tilde{\psi},\label{eq:JumpConditions}
\end{equation}
where $A=2m/a_{1}F(a_{1})$ and $B=m^{2}\beta_{0}a_{1}^{4}/a_{2}^{5}F(a_{2})^{2}$.
Note that $B>0$ in general, and $A>0$ since $F(a_{1})=m-nq(a_{1})=m-nq(0)>0$.
Also, $F(a_{2})=(a_{1}^{2}/a_{2}^{2})(m-nq(a_{2}))$ implies $B=m^{2}\beta_{0}/a_{2}(m-nq(a_{2}))^{2}$.

For this stepfunction modeling, the region outside $r=a_{2}$ satisfies
$\nabla_{\perp}^{2}\tilde{\psi}=0$, by Eqs.~(\ref{eq:OuterRegionEq},\ref{eq:JumpConditions}).
Therefore, this model has the property that if it were modified by
introducing a vacuum in the region $r_{p}<r<r_{w}$ with $r_{p}>a_{2}$,
the equations would be unchanged. To the degree that $j_{z0}(r)$
and $p_{0}(r)$ in Eqs.~(\ref{eq:jz0-specification},\ref{eq:p0-specification})
are very small near $r=r_{w}$, the same conclusions hold for the
numerical full MHD model.

We write the flux $\tilde{\psi}$ as 
\begin{equation}
\tilde{\psi}(r)=\alpha_{1}\psi_{1}(r)+\alpha_{2}\psi_{2}(r)+\alpha_{3}\psi_{3}(r),\label{eq:CulhamExpansion}
\end{equation}
where the basis functions $\psi_{1},\,\psi_{2},\,\psi_{3}$ are described
and computed in the Appendix (see Fig.~10) for this stepfunction
equilibrium. They have $\psi_{1}(0)=0$, $\psi_{1}(r_{t})=1$, $\psi_{1}(r_{w})=0$;
$\psi_{2}(r_{t})=0$, $\psi_{2}(r_{w})=1$, $\psi_{2}(r_{c})=0$;
and $\psi_{3}(r_{w})=0$, $\psi_{3}(r_{c})=1$. There are three conditions
for the three unknowns $\alpha_{1},\,\alpha_{2},\,\alpha_{3}$. The
first is the constant-$\psi$ VR tearing mode jump condition at the
tearing layer at $r=r_{t}$. The second is the resistive thin-wall
jump condition at $r=r_{w}$, and the third is the prescribed feedback
control condition at the control surface $r=r_{c}$:
\begin{equation}
\gamma_{d}\tau_{t}\tilde{\psi}(r_{t})=\left[\tilde{\psi}'\right]_{r_{t}},\label{eq:RPJump}
\end{equation}
\begin{equation}
\gamma\tau_{w}\tilde{\psi}(r_{w})=\left[\tilde{\psi}'\right]_{r_{w}},\label{eq:RWJump}
\end{equation}
\begin{equation}
\tilde{\psi}(r_{c})=-G\tilde{\psi}(r_{w})+K\tilde{\psi}'(r_{w}-).\label{eq:GainEquation}
\end{equation}
Here again $\gamma_{d}$ is the Doppler shifted frequency $\gamma+iku_{z0}=\gamma+i\Omega$;
the plasma velocity enters in only Eq.~(\ref{eq:RPJump}) and we
assume that the velocity shear across the tearing layer is negligible.
Also, $[\cdot]_{r_{t},r_{w}}$ represents the jump in radial derivatives
at $r=r_{t}$ and $r=r_{w}$, respectively. Note that the gain $G$
multiplies the radial (normal) component $\tilde{B}_{r}=im\tilde{\psi}/r$
and $K$ multiplies the poloidal (tangential) normal component $\tilde{B}_{\theta}=-\tilde{\psi}'(r)$.
For sensing of the normal component (for $G$ real), the measured
field consists of the field due to the plasma perturbation as well
as that due to the control coils. This point, which has been discussed
as a reason for preferring tangential sensing\cite{okabayashi}, has
been discussed in Ref.~\cite{FinnChacon1}, where it was shown that,
in cylindrical geometry with idealized coils (i.e.~with a single
poloidal Fourier component), the field due to the plasma alone has
a simple proportionality to the total normal field. Although this
issue is avoided for tangential sensing with $K$ real, it appears
that for $\pi/2$ phase shift ($K$ imaginary) the same considerations
apply.

The results of\textcolor{red}{{} }Ref.~\cite{Sassenberg} show that,
even in a model which contains a second tangential component (here
$\tilde{B}_{z}$), this component is not very important and is zero
if the measurements are made in a vacuum region between the plasma
and the wall. The results in Ref.~\cite{Sassenberg} also show that
in the presence of an inner wall with a much shorter time constant,
this inner wall can be treated as part of the vacuum for small $|\gamma|$,
and that such a simple model with constant-$\psi$ matching and the
thin-wall treatment is qualitatively accurate.

We obtain
\begin{equation}
\gamma_{d}\tau_{t}\alpha_{1}=\Delta_{1}\alpha_{1}+l_{21}\alpha_{2},\label{eq:Basic-1}
\end{equation}
\begin{equation}
\gamma\tau_{w}\alpha_{2}=l_{12}\alpha_{1}+\Delta_{2}\alpha_{2}+l_{32}\alpha_{3},\label{eq:Basic-2}
\end{equation}
and 
\begin{equation}
\alpha_{3}=-G\alpha_{2}+K\left(-l_{12}\alpha_{1}+l_{22}^{(-)}\alpha_{2}\right).\label{eq:Basic-3}
\end{equation}

The quantity $\Delta_{1}$ is the tearing mode matching condition
at $r=r_{t}$ with an ideal wall at $r=r_{w}$, and the entry $\gamma_{d}\tau_{t}$
is based on the VR dispersion relation with the constant-$\psi$ approximation.
(For the resistive-inertial or RI regime, $\gamma_{d}\tau_{t}$ is
replaced by $(\gamma_{d}\tau_{t}')^{5/4}$, but in the presence of
both viscosity and inertia the modes go over to the visco-resistive
regime for $|\gamma_{d}|$ small.) The quantity $\Delta_{2}$ is the
resistive wall matching condition at $r=r_{w}$ with ideal plasma
conditions at $r_{t}$. The inductance coefficients $l_{12}=-\psi_{2}'(r_{w}-),\, l_{21}=\psi_{2}'(r_{t}-),\, l_{32}=\psi_{3}'(r_{w}+)$
as well as $\Delta_{1}=[\psi_{1}']_{r_{t}}$, $\Delta_{2}=[\psi_{2}']_{r_{w}}$,
and $l_{22}^{(-)}=\psi_{2}'(r_{w}-)$ are computed in the Appendix.
See Fig.~10. The pressure affects the values of $\Delta_{1}$, $\Delta_{2}$,
and $l_{22}^{(-)}$ but, since there is no pressure gradient at $r=r_{t}$,
is not included in the tearing layers, where otherwise it could have
stabilizing or destabilizing effects\cite{GlasserGreenejohnson,FinnManheimer}.

Substituting Eq.~(\ref{eq:Basic-3}) into Eqs.~(\ref{eq:Basic-1},\ref{eq:Basic-2})
we obtain
\begin{equation}
\left(\begin{array}{cc}
\Delta_{1}-\gamma_{d}\tau_{t} & \,\,\,\,\,\,\,\,\,\, l_{21}\\
l_{12}-Kl_{32}l_{12} & \,\,\,\,\,\,\,\,\,\,\Delta_{2}-\gamma\tau_{w}-Gl_{32}+Kl_{32}l_{22}^{(-)}
\end{array}\right)\left(\begin{array}{c}
\alpha_{1}\\
\alpha_{2}
\end{array}\right)=0\label{eq:2X2-eigenvalue-primitive}
\end{equation}
or
\begin{equation}
\left(\begin{array}{cc}
\frac{\Delta_{1}}{\tau_{t}}-i\Omega-\gamma & \,\,\,\,\,\,\,\,\,\,\frac{l_{21}}{\tau_{t}}\\
\frac{l_{12}-Kl_{32}l_{12}}{\tau_{w}} & \,\,\,\,\,\,\,\,\,\,\frac{\Delta_{2}-Gl_{32}+Kl_{32}l_{22}^{(-)}}{\tau_{w}}-\gamma
\end{array}\right)\left(\begin{array}{c}
\alpha_{1}\\
\alpha_{2}
\end{array}\right)=0\,\,\,\,\,\,\,\,\,\text{-- or}\,\,\,\,\,\,\,\,\,(\mathsf{A}-\gamma\mathsf{I})\vec{\boldsymbol{\boldsymbol{\alpha}}}=0.\label{eq:2X2-eigenvalueEq}
\end{equation}
The off-diagonal terms couple the resistive plasma ideal wall (rp,iw)
mode and the ideal plasma resistive wall (ip,rw) mode. This leads
to a dispersion relation from $\text{det}(\mathsf{A}-\gamma\mathsf{I})=0$,
or $\gamma^{2}-T\gamma+D=0$, where $T=\text{trace}\mathsf{A}$ and
$D=\text{det}\mathsf{A}$, giving $\gamma=T/2\pm\sqrt{(T/2)^{2}-D}$.
For RI tearing modes rather than VR modes, $\gamma_{d}\tau_{t}$ is
replaced by $(\gamma_{d}\tau_{t}')^{5/4}$. Notice that for $\tau_{t}\rightarrow\alpha\tau_{t}$,
$\tau_{w}\rightarrow\alpha\tau_{w}$ and $\Omega\rightarrow\Omega/\alpha$,
$\gamma$ is replaced by $\gamma/\alpha$, for either the VR or RI
versions. This shows that for $\tau_{t}/\tau_{w}$ fixed marginal
stability is unaffected by changes to $\Omega\tau$, where $\tau=\sqrt{\tau_{t}\tau_{w}}$.

\subsection{Linearized full MHD model}

In this subsection we discuss the full, compressional MHD model used
with smooth current density and pressure profiles. Denoting perturbed
quantities by a tilde, the visco-resistive MHD model reduces to the
following\textcolor{blue}{{} }\textcolor{black}{three} coupled equations:
\begin{equation}
\gamma_{d}\mathbf{\tilde{v}}=\left(\boldsymbol{\nabla}\times\mathbf{\tilde{B}}\right)\times\mathbf{B}_{0}+\mathbf{j}_{0}\times\tilde{\mathbf{B}}-\nabla\tilde{p}+\nu\mathbf{\nabla^{2}}\mathbf{\tilde{v},}\label{eq:2a}
\end{equation}
\begin{equation}
\gamma_{d}\mathbf{\tilde{B}}=\boldsymbol{\nabla}\times\left[\mathbf{\tilde{v}}\times\mathbf{B}_{0}-\eta\boldsymbol{\nabla}\times\mathbf{\tilde{B}}\right],\label{eq:2b}
\end{equation}

\begin{equation}
\gamma_{d}\tilde{p}=-\tilde{\mathbf{v}}\cdot\nabla p_{0}-\Gamma p_{0}\nabla\cdot\tilde{\mathbf{v}},\label{eq:adiabatic}
\end{equation}
where again $\gamma_{d}=\gamma+iku_{z0}=\gamma+i\Omega$, only contributing
a constant Doppler shift due to the uniform toroidal equilibrium flow.
The normalization is such that the (assumed uniform) equilibrium density
is unity; $\Gamma=5/3$ is the adiabatic index. As in Sec.~2, all
perturbations are of the form $e^{i(m\theta+kz-\omega t)}$, where
$\omega=i\gamma$ is the complex frequency and the toroidal mode number
is given by $n=-kR$. For current density and pressure profiles that
are smoothed forms of the stepfunction profiles of Sec.~2, and for
large aspect ratio $R/r_{w}$ so that reduced MHD is fairly accurate,
we obtain results that are in good agreement with those obtained with
the model of Sec.~3.1. These equations are put in dimensionless form
as before, with time in Alfv\'en units using the nominal poloidal
field $B_{\theta0}'(0)r_{w}$ and lengths scaled to $r_{w}$, so that
$B_{\theta0}'(0)r_{w}=r_{w}=1$. The results are reported in terms
of $\beta_{0}=2p_{0}(0)/B_{z0}(0)^{2}$, the Lundquist number $S=\tau_{r}/\tau_{A}$
and the magnetic Prandtl number $Pr=\nu/\eta$. The aspect ratio used
is $R/r_{w}=5$.

In ideal MHD modeling, the modes can be influenced by continuum damping.
We include plasma resistivity, and therefore the continuum is replaced
be discrete damped modes, and collisional transport (represented by
plasma resistivity and viscosity) causes damping in place of the continuum
damping of ideal MHD.

\textcolor{black}{Boundary conditions for the numerical solutions
are applied at the magnetic axis, and the resistive wall with the
control coil and vacuum region coupled into the latter condition.
Since we are interested in modes with $m>1$, the regularity condition
at the magnetic axis at $r=0$ implies that all perturbed quantities
are zero, as opposed to previous RFP studies in Refs.~\cite{RichardsonFinnDelzanno,Sassenberg}
with $m=1$ modes where a more complex regularity condition is needed.
The boundary conditions at the resistive wall are }

\begin{equation}
{\color{black}\gamma_{d}\tilde{B}_{r}(r_{w})=i\mathbf{k}\cdot\mathbf{B}_{0}\tilde{v}_{r}}\label{eq:BC_IdealOhm}
\end{equation}
\begin{equation}
\gamma\tau_{w}\tilde{B}_{r}(r_{w})=[\tilde{B}_{r}^{\prime}]_{r_{w}}\label{eq:RWCondition}
\end{equation}
\begin{equation}
im\tilde{v_{r}}/r+r\partial_{r}(\tilde{v}_{\theta}/r)=0\label{eq:Zero-stress}
\end{equation}
\begin{equation}
ik\tilde{v_{r}}+\partial_{r}\tilde{v_{z}}=0\label{eq:Another-Zero-stress}
\end{equation}
\begin{equation}
\partial_{r}(r\tilde{B}_{\theta})-im\tilde{B_{r}}=0\label{eq:ZeroCurrentDensity}
\end{equation}

\begin{equation}
\partial_{r}\tilde{B_{z}}-ik\tilde{B_{r}}=0\label{eq:AnotherZeroCurrentDensity}
\end{equation}

\begin{equation}
\gamma_{d}\tilde{p}=-\tilde{v}_{r}\partial_{r}p_{0}(r_{w})-\Gamma p_{0}(r_{w})(\nabla\cdot\tilde{\mathbf{v}})_{r_{w}}\label{eq:Adiabatic}
\end{equation}
\begin{equation}
\tilde{B_{r}}(r_{c})=[-(Gr_{w}-K)\tilde{B_{r}}(r_{w})+Kr_{w}\tilde{B_{r}}'(r_{w}-)]/r_{c}.\label{eq:ControlInNumericalBCs}
\end{equation}
The comments made after Eq.~(\ref{eq:GainEquation}) apply as well
to the essentially identical control scheme of Eq.~(\ref{eq:ControlInNumericalBCs}).

\textcolor{black}{The resistive wall and control coil conditions Eqs.~(\ref{eq:RWJump},\ref{eq:GainEquation}),
enter in an analogous way to the reduced MHD model, but take the form
appropriate for the full MHD model in Eqs.~(\ref{eq:RWCondition})
and (\ref{eq:ControlInNumericalBCs}), as discussed in Refs.~\cite{RichardsonFinnDelzanno,Sassenberg}.
No Doppler shift appears in the thin wall boundary condition as this
is in the laboratory frame. In the feedback boundary condition in
Eq.~(\ref{eq:ControlInNumericalBCs}), the tangential component is
$\mathbf{k}\cdot\tilde{\mathbf{B}}=-\tilde{\chi}'$, where the helical
flux $\tilde{\chi}=m\tilde{A}_{z}-kr\tilde{A}_{\theta}$. Using $k\sim\epsilon m/r$,
where $\epsilon=r_{w}/R$, and $\tilde{B}_{z}\sim\epsilon\tilde{B}_{\theta}$,
we find $\mathbf{k}\cdot\tilde{\mathbf{B}}=m\tilde{B}_{\theta}/r+k\tilde{B}_{z}=(m\tilde{B}_{\theta}/r)(1+O(\epsilon^{2}))$.
The control coil equation is coupled into the boundary condition at
$r_{w}$ through a vacuum region solution involving the usual Bessel
function representation for the fields, as discussed in Refs.~\cite{RichardsonFinnDelzanno,Sassenberg}.
The ideal Ohm's law is applied in Eq.~(\ref{eq:BC_IdealOhm}), which
avoids resistive boundary layers near $r_{w}$. }Notice that\textbf{
$\tilde{v}_{r}$} at the wall is allowed, consistent with ideal MHD
(Eq.~(\ref{eq:BC_IdealOhm})) and the finite $\tilde{B}_{r}$ due
to the wall resistivity. Equations (\ref{eq:ZeroCurrentDensity})
and (\ref{eq:AnotherZeroCurrentDensity}) represent the tangential
components of the plasma current being set to zero, consistent with
Eq.~(\ref{eq:BC_IdealOhm}), and preventing an artificial resistive
boundary layer near the wall. (Skin currents in the wall irrelevant.)
\textcolor{black}{The pressure equation is solved in the boundary
condition in Eq.~(\ref{eq:Adiabatic}), giving $\tilde{p}$ very
small (c.f.~Eq.~(\ref{eq:p0-specification})) but finite for $r$
near $r_{w}$. Equations (\ref{eq:Zero-stress}) and (\ref{eq:Another-Zero-stress})
represent a} no-stress boundary condition on $\tilde{\mathbf{v}}$,
reasonable since we are modeling plasmas for which the region near
the wall consists of either cold plasma or vacuum. In general with
the thin wall boundary condition, $\tilde{B}_{r}$ is continuous across
the wall, while the jump in the gradient of $\tilde{B}_{r}$ represents
the current induced in the wall. These boundary conditions are idealized,
and to be sure a more complete treatment of the interaction with the
wall is possible. However, results which we show in the next section
indicate that these boundary conditions do not allow artificial boundary
layers near the walls, producing results that are very similar (in
the numerical modeling) or identical (for the analytic treatment)
to results that would be obtained with a vacuum region just inside
the resistive wall.

\section{Results with zero rotation and gain parameters}

In this section we present results obtained with both the simplified
model, handled analytically as described in Sec.~3.1, and the full
MHD model of Sec.~3.2.

Let us first consider $G=K=0$ and $\Omega=0$ with the simplified
model. Because the $q(r)$ profile is increasing, the negative step
$\Delta j_{z0}$ at $a_{1}<r_{t}$ contributes a destabilizing influence.
The diffuse current density profile in Sec.~2 has a destabilizing
influence for $r<r_{t}$ and a stabilizing influence for $r_{t}<r<r_{w}$.
The negative step $\Delta p_{0}$ is stabilizing for $a_{2}<r_{t}$
or for $a_{2}>r_{t}$ but we assume the latter. In fact, for $\beta_{0}=0$
(or for $a_{2}<r_{t}$) the mode is an \emph{internal} mode, concentrated
in the region $0<r<r_{t}$ and therefore insensitive to the resistive
wall. For $\beta_{0}>0$ the mode is also driven at $r=a_{2}>r_{t}$
and is therefore no longer localized to $r<r_{t}$ and is sensitive
to the resistive wall. In RFPs, i.e.~for decreasing $q(r)$ profiles,
the current density contribution is stabilizing for $r<r_{t}$ and
destabilizing for $r_{t}<r<r_{w}$, so that the mode is not internal,
i.e.~is sensitive to the resistive wall. Thus, the four stability
thresholds in $\lambda=j_{||}/B$ at $r=0$ are distinct and can occur
at zero $\beta$. In a toroidal rather than a cylindrical model for
a tokamak, the modes are more sensitive to the resistive wall because
of poloidal mode coupling. The results in the Appendix show that $\Delta_{1}$
has a destabilizing term due to the current step at $r=a_{1}$, $\sim1/(A+\delta_{1})$,
and one due to the pressure step at $r=a_{2}$, $\sim1/(B+\delta_{2})$;
those results also show that $\Delta_{2}$ has a destabilizing term
from the pressure step only $\sim1/(B+\delta_{2})$. In the Appendix,
we discuss why $\Delta_{1}>\Delta_{2}$, and hence $\beta_{rp,iw}<\beta_{ip,rw}$,
for typical parameters for this model. 

For $\Omega=G=K=0$, we have $T=\text{trace}\mathsf{A}=\Delta_{1}/\tau_{t}+\Delta_{2}/\tau_{w}$
and $D\equiv\text{det}\mathsf{A}=(\Delta_{1}\Delta_{2}-l_{12}l_{21})/\tau_{t}\tau_{w}$.
Also, note that $(T/2)^{2}-D=\left[(\Delta_{1}/\tau_{t}-\Delta_{2}/\tau_{w})^{2}+l_{12}l_{21}\tau_{t}\tau_{w}\right]/4$,
which is nonnegative. (See the Appendix.) As $\beta_{0}$ is increased
from zero, we reach marginal stability $\gamma=0$ at $D\equiv\text{det}\mathsf{A}=0$
or
\begin{equation}
\Delta_{1}=\frac{l_{12}l_{21}}{\Delta_{2}}.\label{eq:beta1}
\end{equation}
This is the resistive plasma-resistive wall limit $\beta=\beta_{rp,rw}$.%
\footnote{We find stability if $1-(a_{1}/r_{c})^{2m}<m-nq(0)$, so if we take
$a_{1}=0.5,\, r_{c}=1.5$, $m=2,\, n=1$ we get $q(0)<1+(a_{1}/r_{c})^{4}=1.01$. %
} Next, we set $\tau_{w}=\infty$ to find, from Eq.~(\ref{eq:2X2-eigenvalueEq}),
$\gamma(\gamma-\Delta_{1}/\tau_{t})=0,$ so that the resistive plasma-ideal
wall stability limit $\beta=\beta_{rp,iw}$ has $\Delta_{1}=0$. Similarly,
setting $\tau_{t}=\infty$ we find that the resistive wall-ideal plasma
limit $\beta=\beta_{ip,rw}$ is at $\Delta_{2}=0$. The condition
$\Delta_{2}<\Delta_{1}$ guarantees that $\beta_{rp,iw}<\beta_{ip,rw}$.
(In Ref.~\cite{finn_rw2} it was concluded that resistive wall modes
could be stabilized by slow rotation for $\beta_{ip,rw}<\beta_{0}<\beta_{rp,iw}$,
the area called Region III in Ref.~\cite{finn_rw2}. This range of
$\beta_{0}$ is empty for the case we consider, with $\beta_{rp,iw}<\beta_{ip,rw}$.)
The analogous ordering for zero-beta reversed field pinches, i.e.~$\lambda_{rp,iw}<\lambda_{ip,rw}$,
holds for all reasonable RFP profiles\cite{RichardsonFinnDelzanno,Sassenberg}.
The ideal wall-ideal plasma limit $\beta_{ip,iw}$ occurs when $\Delta_{1}=\Delta_{2}=\infty$,
both occurring where $B+\delta_{2}\rightarrow0-$, as discussed in
the Appendix. The growth rates for very large $\Delta_{1}$ and\textcolor{blue}{{}
${\normalcolor \Delta}_{2}$ }are not accurate because the constant-$\psi$
approximation and the thin-wall approximation are not accurate there,
but the qualitative behavior is correct and the marginal stability
points are still valid. We choose parameters $a_{1}=0.5,\, a_{2}=0.8,\, r_{w}=1,\, r_{c}=1.5,\, q(0)=0.9$,
and find $r_{t}=0.745$. We summarize in Table I. Notice that, although
$\beta_{ip,rw}<\beta_{ip,iw}$ must hold, the extrapolation process
to $S=\infty$ makes it difficult (as well as unimportant) to obtain
more than two-place accuracy.

\noindent \rule[0.5ex]{1\linewidth}{1pt}

\begin{center}
\begin{tabular}{|c|c|c|c|c|}
\hline 
Model$\downarrow$ ~~~~$\beta_{0}\rightarrow$ & $\beta_{rp,rw}$ & $\beta_{rp,iw}$ & $\beta_{ip,rw}$ & $\beta_{ip,iw}$\tabularnewline
\hline 
\hline 
Analytic & $0.045$ & $0.101$ & $0.383$ & $0.440$\tabularnewline
\hline 
(Analytic) & ($\Delta_{1}=l_{12}l_{21}/\Delta_{2}$) & ($\Delta_{1}=0$) & ($\Delta_{2}=0$) & ($\Delta_{1},\,\Delta_{2}\rightarrow\infty$)\tabularnewline
\hline 
Numerical $S=10^{5}$ & $0.06$ & $0.12$ & $\sim1.5^{*}$ & $\sim1.5^{*}$\tabularnewline
\hline 
\end{tabular}
\par\end{center}

Table 1. Marginally stable $\beta$ values, for the simplified and
numerical models with parameters as in Figs. (1) and (2). Note ({*})
that the two ideal plasma limits are estimated from extrapolations
of the growth rate curves of $S>10^{8}$ to the marginal point in
the ideal MHD regime $S\rightarrow\infty$.

\noindent \rule[0.5ex]{1\linewidth}{1pt}

In Fig.~2a we show the growth rate $\gamma$ in poloidal Alfv\'en
units for $\tau_{w}=10^{3},\,\tau_{t}=10^{4}$, showing the marginal
stability points as in Table 1. We also show $\Delta_{1}$ and $\Delta_{2}$
as functions of $\beta_{0}$. In Fig.~2b we show $\gamma\tau_{A}$
vs $\beta_{0}$ for the full MHD model. The value $\beta_{rp,iw}$
is found by setting $\tau_{w}$ very large; $\beta_{ip,rw}$ is found
by a convergence study for large Lundquist number $S$.

In Fig.~3 we include feedback (real $G$, $K$) but with $\Omega=G_{i}=K_{i}=0$,
and show the stability diagram for four values of $\beta_{0}$, both
for the simplified model and the full MHD model. Note that $G=K=0$
is in the stable region for the lowest value of $\beta_{0}$ in Fig.~3a,
consistent with ${\normalcolor {\color{red}{\normalcolor \beta_{0}<\beta_{rp,rw}}}}$.
Also, the results are consistent with the top line becoming vertical
for $\beta_{0}=\beta_{rp,iw}=0.101$, and that the slope of the top
line approaches that of the bottom line as $\beta\rightarrow\beta_{ip,iw}$,
where $\Delta_{1},\,\Delta_{2}\rightarrow\infty$. The results in
Fig.~3b and c, with the full MHD model, show similar results. It
is thus possible to stabilize the tearing mode above $\beta_{rp,iw}$
and, for the simplified model, technically up to $\beta_{ip,iw}$.
In Ref.~\cite{Finn2006} a stable window was shown to exist up to
$\beta_{ip,iw}$, as in the present results; in Refs.~\cite{RichardsonFinnDelzanno,Sassenberg},
with finite viscosity, the limit was slightly below $\beta_{ip,iw}$.
Indeed, stability ($\text{Re}(\gamma)<0$) is guaranteed for the simplified
model if the trace in Eq.~(\ref{eq:2X2-eigenvalueEq}) is negative
and the determinant is positive. The trace condition for stability
with feedback, $T<0$, gives
\begin{equation}
G>Kl_{22}^{(-)}+\frac{\Delta_{2}}{l_{32}}+\frac{\Delta_{1}}{l_{32}}\frac{\tau_{w}}{\tau_{t}},\label{eq:TraceCondition}
\end{equation}
and depends on $\tau_{w}/\tau_{t}$\cite{Finn2006}. The determinant
condition $D>0$ is independent of $\tau_{w}/\tau_{t}$ and gives
\begin{equation}
\Delta_{1}G<\Delta_{1}Kl_{22}^{(-)}+\frac{\Delta_{1}\Delta_{2}}{l_{32}}+\frac{l_{12}l_{21}}{l_{32}}\left(Kl_{32}-1\right).\label{eq:DeterminantCondition}
\end{equation}
(Recall that $\Delta_{1}$ can have either sign; this inequality is
valid for either sign of $\Delta_{1}$.) The upper and lower straight
lines correspond to the determinant condition in Eq.~(\ref{eq:DeterminantCondition})
and the trace condition in Eq.~(\ref{eq:TraceCondition}), respectively.
The upper line (independent of $\tau_{w}/\tau_{t}$) is the marginal
stability curve for the purely growing tearing mode. The lower line
(with intercept depending on $\tau_{w}/\tau_{t}$) corresponds to
a complex root driven unstable by the feedback below the line. See
Refs.~\cite{Finn2006,RichardsonFinnDelzanno,Sassenberg}. Notice
that the determinant condition indeed gives a vertical line at $\beta_{rp,iw}$,
where $\Delta_{1}=0$, and it is also clear that the slopes of the
lines become equal for $\Delta_{1}$ large, so that the stable region
disappears as $\beta\rightarrow\beta_{ip,iw}$.

If we look for the intersection of the $T=0$ line and the $D=0$
line we find
\begin{equation}
\Delta_{1}^{2}=\frac{\tau_{t}}{\tau_{w}}l_{12}l_{21}\left(Kl_{32}-1\right).\label{eq:IntersectionPoint}
\end{equation}
That is, at this intersection we must have $l_{32}K>1$. Note that
this implies that the coupling coefficient $\sim a_{21}$ in Eq.~(\ref{eq:2X2-eigenvalueEq})
is negative in the stable region. Also, this holds regardless of the
sign of $\Delta_{1}$, i.e.~with $\beta_{0}$ on either side of $\beta_{rp,iw}$
and for $\beta_{0}=\beta_{rp,iw}$ ($\Delta_{1}=0$) this intersection
has $K=1/l_{32}$. More importantly, this intersection occurs for
rapidly increasing $K$ (and $G$) as $\Delta_{1}$ increases. The
slopes in $(G,K)$ of the marginal stability lines, from Eqs.~(\ref{eq:TraceCondition},\ref{eq:DeterminantCondition}),
approach each other rapidly as $\Delta_{1}$ increases, so that, although
the theoretical limit for feedback stabilization is $\beta_{ip,iw}$,
the practical limit is a few times $\beta_{rp,iw}$. This practical
limit can be below or above $\beta_{ip,rw}$. 

Another point is that, whereas the two straight lines intersect at
a point for the simplified model, the lower stability boundary in
Fig.~3(c) develops curvature for the full MHD model. This curvature
is reproduced qualitatively by using the RI version of the simplified
model, with $\gamma_{d}\tau_{t}\rightarrow(\gamma_{d}\tau_{t}')^{5/4}$

The major conclusions of this section are that feedback stabilization
appears to be practically possible well above $\beta_{rp,iw}$ and
possibly above $\beta_{ip,rw}$. Further, the simplified model captures
well the qualitative behavior of the full MHD model. We also note
that in a toroidal configuration at moderate aspect ratio the ideal
plasma limits will be much lower, as the toroidicity affects the stability
at the same order as pressure and current, and thus the stable regions
will more easily approach $\beta_{ip,iw}$.

\section{Results with plasma rotation and complex gain $G_{i}$}

In this section we show analytic and numerical results for the appropriate
equilibria, with plasma rotation and complex gain $G_{i}$, both for
the simplified model and the full MHD model.

Figure 4, for $\beta_{0}<\beta_{rp,iw}$, shows that the stable region
increases in size with $\Omega$ ($G_{i}=K_{i}=0$) in this range
for the simplified model. ($\Omega\rightarrow-\Omega$ gives identical
results for the growth rate, with $\gamma\rightarrow\gamma^{*}$,
both for the simplified model and for the full MHD model.) This stabilization
is expected because the mode in this regime is a resistive wall tearing
mode, and plasma rotation relative to the wall stabilizes by supressing
flux from penetrating the wall. A close look at the numerical results
in Fig.~4b shows that for low rotation, for $0<\Omega<0.001$, the
stable region actually shrinks along some sections of the marginal
stability curve. This is related to the fact that in the RI regime
low rotation initially destabilizes resistive wall modes, followed
by stabilization for higher rotation. This behavior is explained by
the mode-coupling picture of Ref.~\cite{finn_rw4} and is even more
noticeable for ideal plasma resistive wall modes\cite{betti-freidberg,finn_rw4}.
Larger $\Omega$ is stabilizing for the full MHD model and the expanding
stable region develops a tail toward negative $G$ and $K$. This
general behavior of stabilization as $\Omega$ increases is consistent
with the observation in Ref.~\cite{Finn2006} that increasing $\tau_{w}/\tau_{t}$
is stabilizing in this regime. 

The curvature seen in Fig.~4b with $\Omega=0$ at the tip is seen
all along the curve to the right. This is consistent with the fact
that the RI model is reasonable for this curve because marginal stability
there has real frequency; and in the RI regime the $(\gamma_{d}\tau_{t}')^{5/4}$
term does indeed cause curvature (not shown.) On the upper (left)
curve, marginal stability has $\gamma=0$, so that the VR dispersion
relation is correct there, giving a linear marginal stability curve. 

Figure 5 shows a case with $\beta_{0}>\beta_{rp,iw}$, in which the
stable area is observed to \emph{decrease} as the plasma rotation
$\Omega$ increases, for both the simplified model and full MHD. The
explanation is as follows: In this regime the tearing mode is unstable
even with an ideal wall, so \emph{lower} $\Omega$ allows the feedback
flux to penetrate the resistive wall faster. These results can be
interpreted in terms of a virtual wall inside $r=r_{w}$ for the upper
curve, but not the lower curve, which has complex frequency even for
$\Omega=0$. As discussed in Ref.~\cite{FinnChacon1}, there is an
equivalence between $G_{i}$ and wall rotation in the presence of
a single value of $k_{z}=k$. (So this equivalence is not exact in
nonlinear theory.)  This is evident in Eq.~(\ref{eq:2X2-eigenvalueEq}):
the effective wall rotation rate $\Omega_{w}=ku_{zw}$ is given by
\begin{equation}
\Omega_{w}=\frac{l_{32}G_{i}}{\tau_{w}}.\label{eq:RotationGiEquivalence}
\end{equation}
Results (not shown) with $G_{i}$ such that the equivalent wall rotation
$\Omega_{w}$ is equal to the values of the plasma rotation in Fig.~5
give identical results.

Figure 6 shows a case, again with $\beta_{0}>\beta_{rp,iw}$ and the
same parameters but with $\Omega=0.005,\, K_{i}=0$ and four values
of $G_{i}$. Note however, that in Fig.~6 both the analytic and numerical
models have $\tau_{w}=2\times10^{4}$. In the configuration of Fig.~1(a)
we calculate $l_{32}=2.2$. The value of $G_{i}$ corresponding to
$\Omega_{w}=\Omega$ is $G_{iw}=\Omega_{w}\tau_{w}/l_{32}$ in the
analytic model, showing that the optimal value of $G_{i}$ is where
the relative rotation rate vanishes, $\Omega-\Omega_{w}=0$, at $G_{i}=45$.
Also, the stability regions are symmetric about $G_{i}=G_{iw}$: In
the plasma frame $G_{i}$ enters in Eq.~(\ref{eq:2X2-eigenvalue-primitive})
as\textcolor{black}{{} ${\normalcolor \Omega\tau_{w}-l_{32}G_{i}=l_{32}(G_{iw}-G_{i}){\normalcolor =(\Omega-\Omega_{w})\tau_{w}}}$
}and $\gamma\rightarrow\gamma^{*}$ shows that $\gamma_{real}$ is
an even function of $G_{iw}-G_{i}$. Similar behavior is seen for
the full MHD model, with optimal $G_{i}\approx40$. Indeed, the boundary
conditions related to the resistive wall and feedback (Sec.~3.2)
also show this equivalence between rotation and $G_{i}$. 

As expected, it is clear from this discussion that there is some advantage
in having two resistive walls, with complex gain to give effective
rotation to the outer wall\cite{gimblettRotatingSecondWall,Fitzpatrick-Jensen,FinnChacon1},
in the regime $\beta_{0}<\beta_{rp,iw}$. However, there is no such
advantage in the regime $\beta_{0}>\beta_{rp,iw}$, since optimal
control in this latter regime allows the flux from the outside to
penetrate the wall to get into the plasma.

As in the previous section, we conclude that the simplified model
captures the essential physics of the full MHD model.

\section{Studies with plasma rotation and complex gain $K_{i}$}

In this section we show results with $G_{i}=0$ but with imaginary
gain $K_{i}$, both for the simplified model and the full MHD model.

The effect of $K_{i}$, the imaginary part of the tangential gain
$K$, on the results is not as transparent as that of $G_{i}$ because
$K$ occurs in two matrix elements in Eq.~(\ref{eq:2X2-eigenvalueEq}).
In Fig.~7a we show results using the simplified model with parameters
as in Fig.~2, with $\beta<\beta_{rp,iw}$, $\Omega=0$ and four values
of $K_{i}$. In Fig.~7b we show corresponding results with $\beta_{0}>\beta_{rp,iw}$.
Symmetry about $K_{i}=0$ is apparent in both and is easily proved
by arguments like those in the previous section. As with $\Omega$
and $G_{i}$, increasing $|K_{i}|$ shrinks the stable region for
$\beta_{0}>\beta_{rp,iw}$. However, we observe that the stable region
also shrinks with increasing $|K_{i}|$ for $\beta_{0}<\beta_{rp,iw}$,
indicating that the behavior with respect to $K_{i}$ differs significantly
from the behavior with varying $\Omega$.

Results with finite $\Omega$ and $K_{i}$, for $\beta_{0}<\beta_{rp,iw}$
are shown in Fig.~8 for the simplified and full MHD models. Here
we see a  difference between the simplified and reduced MHD models.
In this range of $\beta_{0}$, for the simplified model the stable
region is largest near ${\color{magenta}{\normalcolor K_{i}=0}}$,
and returns the result to near that of $\Omega=0$ in Fig.~4(a) for
$K_{i}=-3$, but decreases the stable region outside of these values.
However, no symmetry about the optimal value is observed. In Fig.~8(b)
the numerical results with finite $\Omega$ and $K_{i}$ are shown,
where the optimal $K_{i,opt}\approx1$ and the stable region decreases
in size more slowly for $K_{i}<0$ than for $K_{i}>0$, in qualitative
agreement with the simplified model. The optimal $K_{i}$ can appear
on either side of $K_{i}=0$ here, as the effects of wall time and
plasma response compete, but the stable regions tend to be more prominent
for $K_{i}<0$ for $\Omega>0$.%
\footnote{Indeed, an argument along the lines of that in the previous section
shows that $\gamma_{real}$ is a symmetric function of $\Omega\tau_{w}+l_{32}l_{22}^{(-)}K_{i}$
and $K_{i}$, the first dependence coming through the $a_{22}$ matrix
element and the second dependence from the $a_{21}$ matrix element.
But such a function is not symmetric in $K_{i}$ with $\Omega$ held
fixed.%
}

Results with $\beta_{0}>\beta_{rp,iw}$ and finite $\Omega$ and $K_{i}$
are shown in Fig.~9(a) for the simplified model. In the results shown
in Fig.~9a, the stable region is largest for $K_{i}=-1$ but shrinks
as $K_{i}$ changes away from this optimal value. The result is not
symmetric\textcolor{blue}{,} as can be seen by the similarity between
the results for $K_{i}=-4$ and $K_{i}=0$. Results in Fig.~9(b)
for the full MHD model have some similar aspects, but differ in that
the width of the stable region increases with $K$ as the boundary
becomes curved with increasing $K_{i}$. Again, no symmetry about
any value of $K_{i}$ is observed. These results show that large $K_{i}$
(positive or negative) destabilize, but for moderate $K_{i}$ with
rotation, the full MHD results vary significantly from those of reduced
MHD. It is in general true that with rotation the stable region has
some optimal $K_{i}$, but in full MHD it is not the same shape as
$\Omega=K_{i}=0$. It can in fact be larger in some cases. In contrast
to previous sections, we observe that the results using the full MHD
model are captured by the simplified model in a broad sense, but some
differences are observed in detail. 

Though not shown here, in highly limited regions of parameter space
as the stable regions approach marginality, weakly growing modes can
appear within and distort the stable regions in the full MHD description.
Likewise, isolated regions of stability can appear in the unstable
region near marginality, rapidly moving to negative $G$ and $K$
as the original stable region moves to positive $G$ and $K$. These
behaviors in marginally stable regions of parameter space are beyond
the scope of this paper, but will be considered in context as we next
look to investigate analogous systems in toroidal geometry.

\section{Summary and conclusions}

In this paper we have used a cylindrical linear model for a tokamak
to make initial investigations in tokamak geometry into feedback control
using complex gains $G$ and $K$, multiplying the measured radial
and poloidal magnetic field components, respectively, in the presence
of plasma resistivity and rotation. This model has four stability
thresholds in the following order: $\beta_{rp,rw}<\beta_{rp,iw}<\beta_{ip,rw}<\beta_{ip,iw}$,
where $rp$ and $ip$ represent resistive plasma and ideal plasma,
respectively, and $rw$ and $iw$ stand for resistive wall and ideal
wall. We have determined the region of stability as a function of
the real parts of the gains $G$ and $K$. For $\beta_{0}<\beta_{rp,iw}$,
rotation $\Omega$ or imaginary gain $G_{i}$, which is equivalent
to rotation of the resistive wall\cite{FinnChacon1}, stabilizes.
This is because in this regime, the tearing mode is unstable with
a resistive wall but not with an ideal wall, and rotation can easily
stabilize resistive wall tearing modes\cite{finn_rw1}. In this regime,
$K_{i}$ is actually destabilizing and is therefore not equivalent
to rotation. For $\beta_{0}>\beta_{rp,iw}$, on the other hand, plasma
rotation $\Omega$ and $G_{i}$ are both destabilizing, while results
for $K_{i}$ are more complex. Above $\beta_{rp,iw}$ and for nonzero
plasma rotation $\Omega$, the optimal value for $G_{i}$ is the value
for which the equivalent wall rotation equals the plasma rotation,
and for this value of $G_{i}$ stability is possible well above $\beta_{rp,iw}$,
as for $\Omega=G_{i}=0$. There is also an optimum value of $K_{i}$
in both ranges of $\beta_{0}$, but its value and shape in $G,K$
space cannot easily be determined by a simple equivalence with rotation,
indeed the situation for $\beta_{0}>\beta_{rp,iw}$, shown in Fig.~9(b),
is more complex than for $\beta_{0}<\beta_{rp,iw}$. These results
have been found by both analysis on a reduced resistive MHD model
with simple stepfunction current density and pressure profiles and
a general MHD model with smooth profiles; the results and conclusions
from both models are very similar.

The fact that rotation or $G_{i}$ is stabilizing for $\beta_{0}<\beta_{rp,iw}$
and destabilizing for $\beta_{0}>\beta_{rp,iw}$ suggests the importance
of modeling resistive wall modes and their control including plasma
resistivity, at least for resonant modes. The use of ideal MHD modeling
with a resistive wall tacitly assumes that $\beta_{rp,rw}\lessapprox\beta_{ip,rw}$
and $\beta_{rp,iw}\lessapprox\beta_{ip,iw}$, which is not consistent
with the results from our cylindrical model, namely $\beta_{rp,rw}<\beta_{rp,iw}<\beta_{ip,rw}<\beta_{ip,iw}$.
If the latter ordering holds in toroidal geometry, then modeling using
non-ideal MHD for resistive wall modes in toroidal geometry is also
important.

\section*{Acknowledgments}

The work of D.~P.~Brennan was supported by the DOE Office of Science,
Fusion Energy Sciences under Contract No DE-SC0004125. The work of
J.~M.~Finn was supported by the DOE Office of Science, Fusion Energy
Sciences and performed under the auspices of the NNSA of the U.S.
DOE by LANL, operated by LANS LLC under Contract No DEAC52- 06NA25396.

\section*{Appendix. Calculations for stepfunction model}

In this appendix we show the steps necessary to compute $\psi_{1},\,\psi_{2},$
and $\psi_{3}$, i.e.~the quantities $l_{12},\, l_{21},\, l_{32},\,\Delta_{1},\,\Delta_{2},$
and $l_{22}^{(-)}$. We first define auxiliary functions $\phi_{1},\,\phi_{t},\,\phi_{2},\,\phi_{w},$
and $\phi_{c}$ with $\phi_{1}(0)=0,\,\phi_{1}(a_{1})=1,\,\phi_{1}(r_{t})=0$.
The four radii are $a_{1}$, where the current density step is; $r_{t}$,
the tearing layer; $a_{2}$, where the pressure step is; $r_{w}$,
the radius of the resistive wall; and $r_{c}$, the position of the
control surface. The other three functions $\phi_{t},\,\phi_{2},\,\phi_{w}$
are defined similarly. We have
\[
\phi_{1}(r)=(r/a_{1})^{m}\,\,\,\text{for}\,\, r<a_{1}\,\,\,\text{and}\,\,\,
\]
\[
\phi_{1}(r)=\frac{(r_{t}/r)^{m}-(r/r_{t})^{m}}{(r_{t}/a_{1})^{m}-(a_{1}/r_{t})^{m}}\,\,\text{for}\,\,\, a_{1}<r<r_{t}.
\]
Similar expressions hold for $\phi_{t},\dots,\phi_{c}$. See Fig.~10.
We find 
\[
\phi_{1}'(a_{1}-)=\frac{m}{a_{1}};\,\,\,\phi_{1}'(a_{1}+)=-\frac{m}{a_{1}}\frac{(r_{t}/a_{1})^{m}+(a_{1}/r_{t})^{m}}{(r_{t}/a_{1})^{m}-(a_{1}/r_{t})^{m}}.
\]
This leads to
\[
\delta_{1}\equiv[\phi_{1}']_{a_{1}}=-\frac{2m}{a_{1}}\frac{(r_{t}/a_{1})^{m}}{(r_{t}/a_{1})^{m}-(a_{1}/r_{t})^{m}}.
\]
All other quantities are computed in the same manner:
\[
k_{t1}=\phi_{t}'(a_{1}+)=\frac{2m}{a_{1}}\frac{1}{(r_{t}/a_{1})^{m}-(a_{1}/r_{t})^{m}},
\]
\[
k_{1t}=-\phi_{1}'(r_{t}-)=\frac{2m}{r_{t}}\frac{1}{(r_{t}/a_{1})^{m}-(a_{1}/r_{t})^{m}},
\]
\[
k_{2t}=\phi_{2}'(r_{t}+)=\frac{2m}{r_{t}}\frac{1}{(a_{2}/r_{t})^{m}-(r_{t}/a_{2})^{m}},
\]
\[
k_{t2}=-\phi_{t}(a_{2}-)=\frac{2m}{a_{2}}\frac{1}{(a_{2}/r_{t})^{m}-(r_{t}/a_{2})^{m}},
\]
\[
k_{w2}=\phi_{w}'(a_{2}+)=\frac{2m}{a_{2}}\frac{1}{(r_{w}/a_{2})^{m}-(a_{2}/r_{w})^{m}},
\]
\[
k_{2w}=-\phi_{2}'(r_{w}-)=\frac{2m}{r_{w}}\frac{1}{(r_{w}/a_{2})^{m}-(a_{2}/r_{w})^{m}},
\]
\[
k_{ww}^{(-)}=-\phi_{w}'(r_{w}-)=\frac{m}{r_{w}}\frac{(r_{w}/a_{2})^{m}+(a_{2}/r_{w})^{m}}{(r_{w}/a_{2})^{m}-(a_{2}/r_{w})^{m}},
\]
\[
k_{cw}=\phi_{c}'(r_{w}+)=\frac{2m}{r_{w}}\frac{1}{(r_{c}/r_{w})^{m}-(r_{w}/r_{c})^{m}},
\]
\[
\delta_{t}=[\phi_{t}']_{r_{t}}=-\frac{m}{r_{t}}\left[\frac{(r_{t}/a_{1})^{m}+(a_{1}/r_{t})^{m}}{(r_{t}/a_{1})^{m}-(a_{1}/r_{t})^{m}}+\frac{(a_{2}/r_{t})^{m}+(r_{t}/a_{2})^{m}}{(a_{2}/r_{t})^{m}-(r_{t}/a_{2})^{m}}\right],
\]
\[
\delta_{2}=[\phi_{2}]_{a_{2}}=-\frac{m}{a_{2}}\left[\frac{(a_{2}/r_{t})^{m}+(r_{t}/a_{2})^{m}}{(a_{2}/r_{t})^{m}-(r_{t}/a_{2})^{m}}+\frac{(r_{w}/a_{2})^{m}+(a_{2}/r_{w})^{m}}{(r_{w}/a_{2})^{m}-(a_{2}/r_{w})^{m}}\right],
\]
and 
\[
\delta_{w}=[\phi_{w}']_{r_{w}}=-\frac{m}{r_{w}}\left[\frac{(r_{w}/a_{2})^{m}+(a_{2}/r_{w})^{m}}{(r_{w}/a_{2})^{m}-(a_{2}/r_{w})^{m}}+\frac{(r_{c}/r_{w})^{m}+(r_{w}/r_{c})^{m}}{(r_{c}/r_{w})^{m}-(r_{w}/r_{c})^{m}}\right].
\]

We set $\psi(r)=a_{1}\phi_{1}(r)+a_{t}\phi_{t}(r)+a_{2}\phi_{2}(r)$.
The condition $\psi_{1}(a_{1})=1$ implies $\alpha_{t}=1$, and Eq.~(\ref{eq:JumpConditions})
implies $[\psi_{1}']_{a_{1}}=-A$ and $[\psi_{2}']_{a_{2}}=-B$. From
these we find $\delta_{1}a_{1}+k_{t1}a_{t}=-Aa_{1}$ and $\delta_{2}a_{2}+k_{t2}a_{t}=-Ba_{2}$
or
\[
a_{1}=-\frac{k_{t1}}{A+\delta_{1}},\,\,\,\,\, a_{2}=-\frac{k_{t2}}{B+\delta_{2}}.
\]
We conclude 
\[
\Delta_{1}=\delta_{t}-\frac{k_{1t}k_{t1}}{A+\delta_{1}}-\frac{k_{2t}k_{t2}}{B+\delta_{2}}.
\]
Similar calculations, plus the fact that $\phi_{c}=\psi_{3}$ show
\[
l_{12}=-\frac{k_{t2}k_{2w}}{B+\delta_{2}},
\]
\[
\Delta_{2}=\delta_{w}-\frac{k_{w2}k_{2w}}{B+\delta_{2}},
\]
\[
l_{22}^{(-)}=\frac{k_{w2}k_{2w}}{B+\delta_{2}}+k_{ww}^{(-)},
\]
\[
l_{21}=-\frac{k_{w2}k_{2t}}{B+\delta_{2}},
\]
and 
\[
l_{32}=k_{cw}.
\]
 A sketch of $\psi_{1}-\psi_{3}$, as well as $\phi_{1},\phi_{t},\phi_{2},\phi_{w}$,
and $\phi_{c}$ is shown in Fig.~10.

The terms proportional to $1/(A+\delta_{1})$ are due to the destabilizing
influence of the current density gradient at $a_{1}$. Those proportional
to $1/(B+\delta_{2})$ are due to the destabilizing influence of the
pressure gradient at $a_{2}$. The condition $\Delta_{1}>\Delta_{2}$
gives
\[
\Delta_{1}-\Delta_{2}=\delta_{t}-\delta_{w}-\frac{k_{1t}k_{t1}}{A+\delta_{1}}.
\]
The term $\delta_{t}-\delta_{w}$ depends only one the geometry, i.e.~on
$a_{1},\, r_{t},\, a_{2},\, r_{w}$, and $r_{c}$. It is positive
if $r_{w}-a_{2}$ or $r_{c}-r_{w}$ is small enough, which we assume.
The term $-k_{1t}k_{t1}/(A+\delta_{1})$, from the drive by the current
gradient inside $r_{t}$, is positive for $A\sim\Delta j_{z0}$ small
and goes to infinity as $A+\delta_{1}\rightarrow0-$. We consider
cases in which the drive due to the current, while not sufficient
to drive the instability for zero pressure, is fairly large, so that
$\beta_{rp,rw}$ and $\beta_{rp,iw}$ are small. In the simplified
model, the values $\beta_{ip,rw}$ and $\beta_{ip,iw}$ are fairly
large (and those values for the numerical model are large) because
for an ideal plasma $\tilde{\psi}(r_{t})$ is zero, and therefore
any unstable mode must be driven solely by the pressure gradient in
the region $r_{t}<r<r_{w}$. Summarizing, for the geometry and and
profiles we consider, $\Delta_{1}-\Delta_{2}$ should be positive,
which implies $\beta_{rp,iw}<\beta_{ip,rw}$. Poloidal mode coupling
in a torus, $m\rightarrow m\pm1$, prevent the shielding of the mode
inside $r=r_{t}$ from the region for $r>r_{t}$. This will be the
subject of a future publication.

Notice that $l_{12},\, l_{21}$ are positive for $B\rightarrow0$
(pressure $p_{0}\rightarrow0$) and go to infinity as $B+\delta_{2}\rightarrow0-$.
Also, $\Delta_{1},\,\Delta_{2}\rightarrow+\infty$ as $B+\delta_{2}\rightarrow0-$,
and $l_{22}^{(-)}\rightarrow-\infty$ in this limit. As we shall discuss
in Sec.~III, the limit $B+\delta_{2}\rightarrow0-$, where $\Delta_{1},\,\Delta_{2}\rightarrow+\infty$,
is the ideal plasma-ideal wall limit $\beta_{ip,iw}$.

These quantities are used in the dispersion relation in Eq.~(\ref{eq:2X2-eigenvalueEq})
to obtain the results in Sec.~3-6.

\bibliographystyle{plain}
\bibliography{BrennanFinn}

\newpage{}
\begin{figure}
\caption{\protect\includegraphics[scale=0.65]{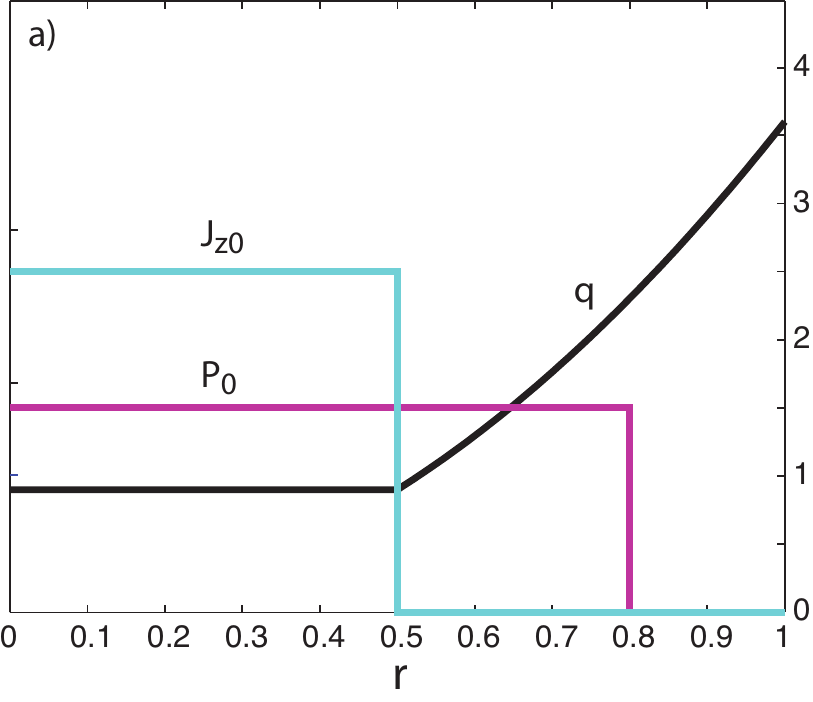}\protect\includegraphics[scale=0.65]{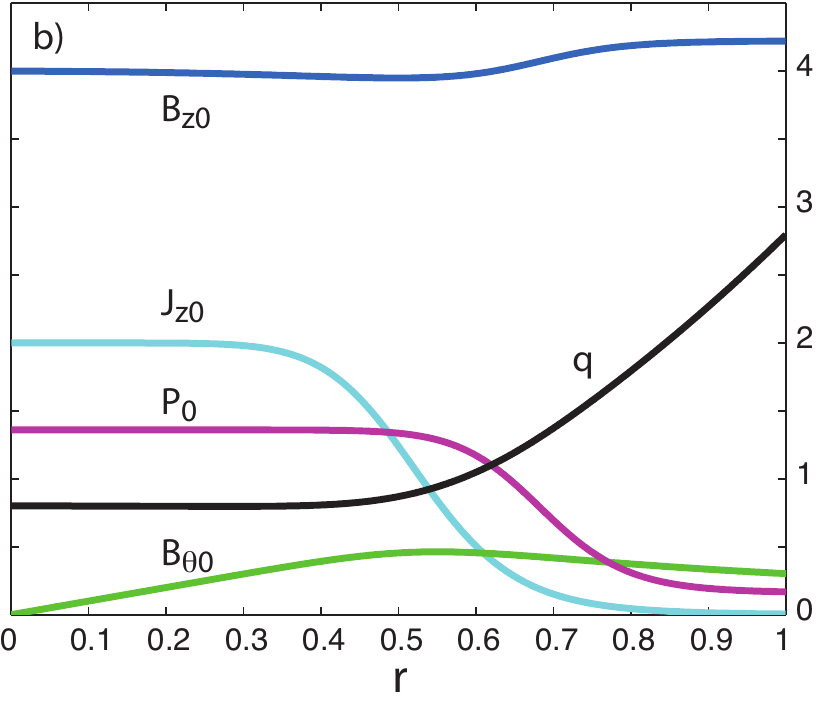}}

\textcolor{black}{Equilibrium current density $j_{z0}(r),$ pressure
$p_{0}(r)$ and safety factor $q(r)$, showing $q=m/n=2/1$, for (a)
the stepfunction model used in the analytic studies with $a_{1}=0.5,\, a_{2}=0.8,\, r_{w}=1,\, r_{c}=1.5,\, q(0)=0.9$,
and (b) the smooth model used in the numerical studies with $a_{1}=0.55,\, a_{2}=0.7,\, r_{w}=1,\, r_{c}=1.5,\, q(0)=0.8$.}
\end{figure}
\begin{figure}
\caption{\protect\includegraphics[scale=0.3]{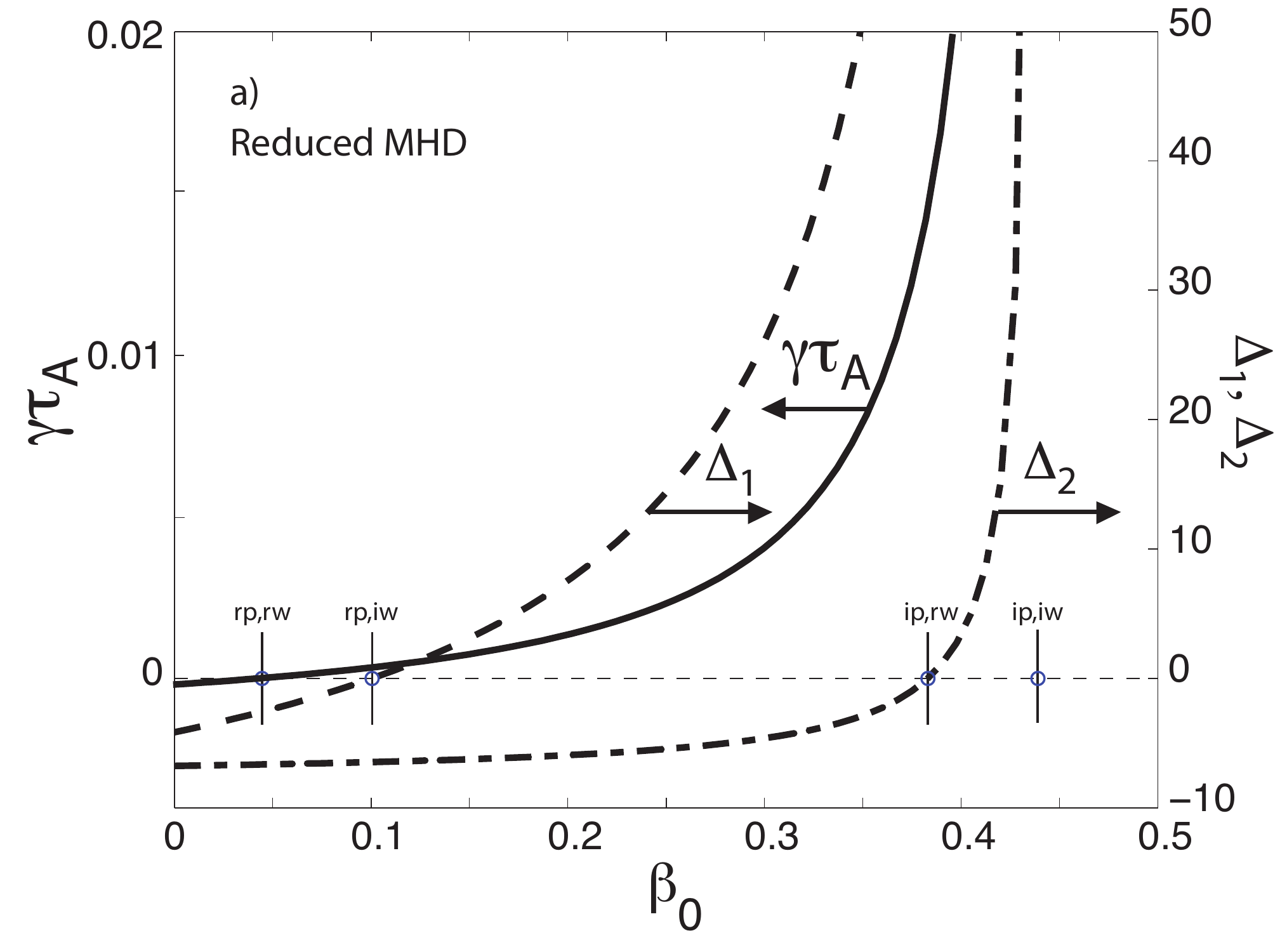}\protect\includegraphics[scale=0.325]{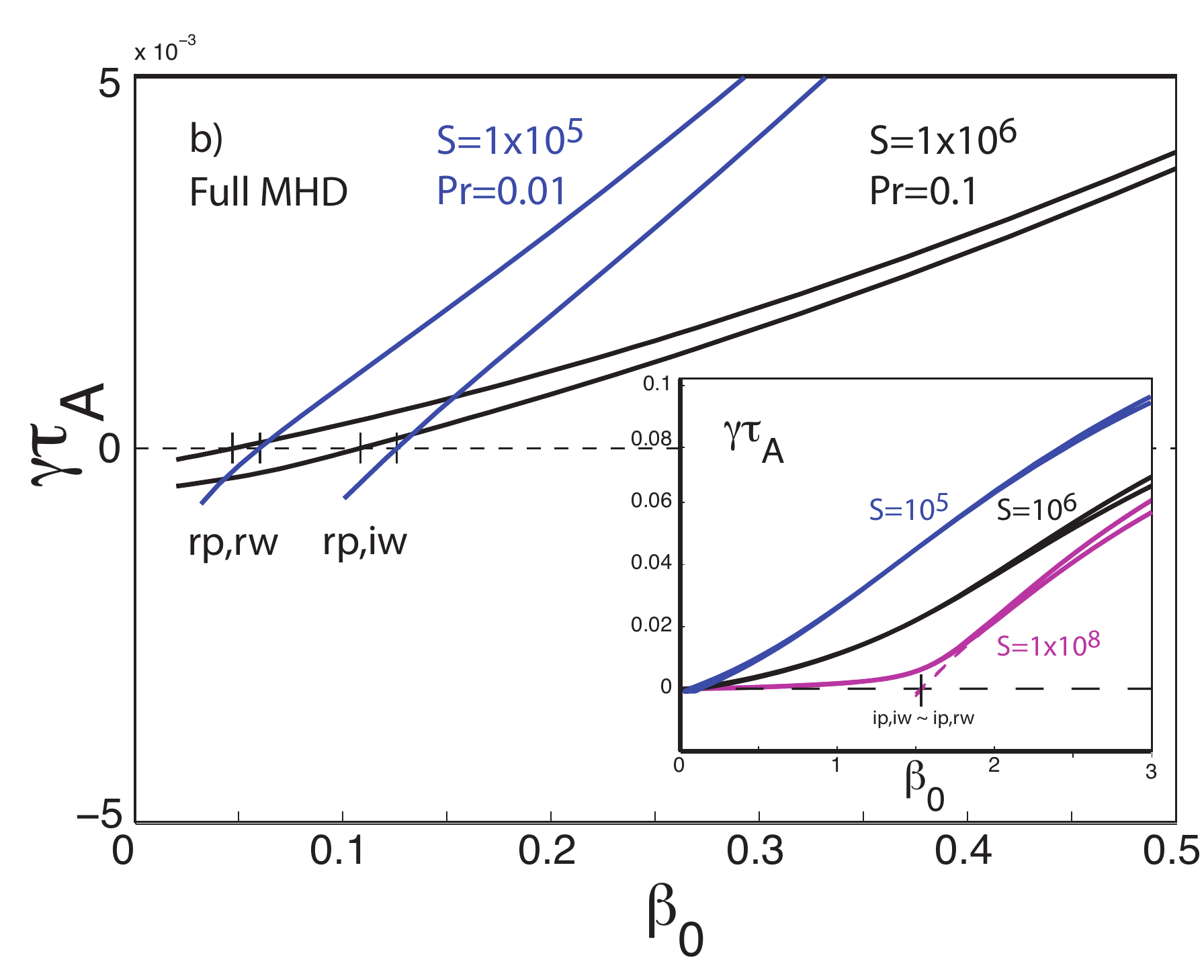}}

\textcolor{black}{Growth rate $\gamma\tau_{A}$ as a function of $\beta_{0}=2p_{00}/B_{0}^{2}$
for (a) the reduced MHD - stepfunction analytic model with parameters
$\tau_{w}=10^{3},\,\tau_{t}=10^{4}$, and (b) the full MHD model with
$S=\tau_{R}/\tau_{A}=10^{5}$, $10^{6}$ and $10^{8}$ with both a
highly conducting $\tau_{w}=10^{10}$ and transparent $\tau_{w}=0$
wall. In (a) we also show jump quantities $\Delta_{1},\,\Delta_{2}$
as functions of $\beta_{0}$. The four $\beta_{0}$ limits $\beta_{rp,rw}<\beta_{rp,iw}<\beta_{ip,rw}<\beta_{ip,iw}$
marked in (a) are given in Table 1. In (b) the lower limits $\beta_{rp,rw}<\beta_{rp,iw}$
are indicated and change slightly with $S$. The extrapolation to
$S=\infty$ is indicated in the inset.}
\end{figure}
\begin{figure}
\caption{\protect\includegraphics[scale=0.3]{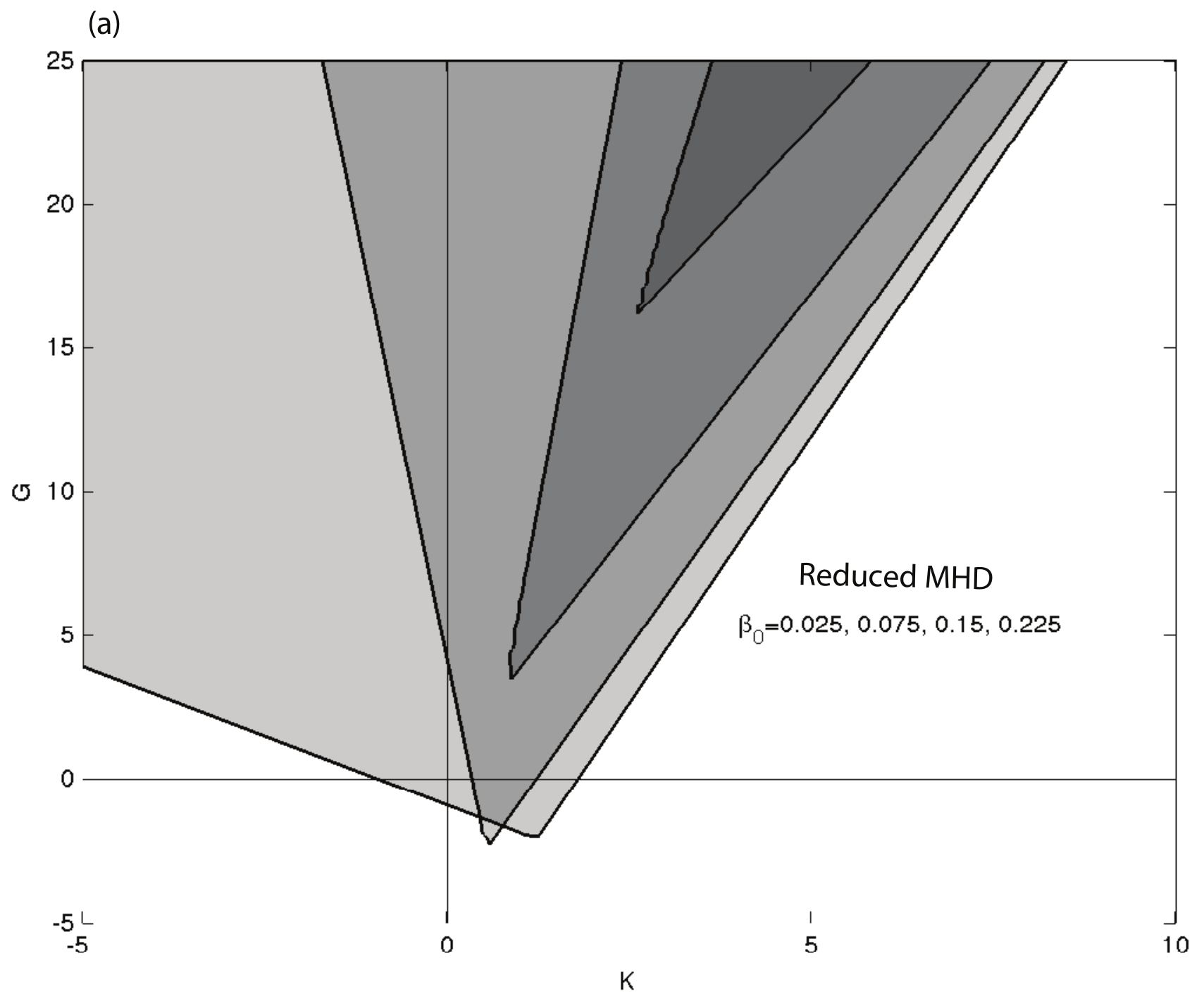}\protect\includegraphics[scale=0.3]{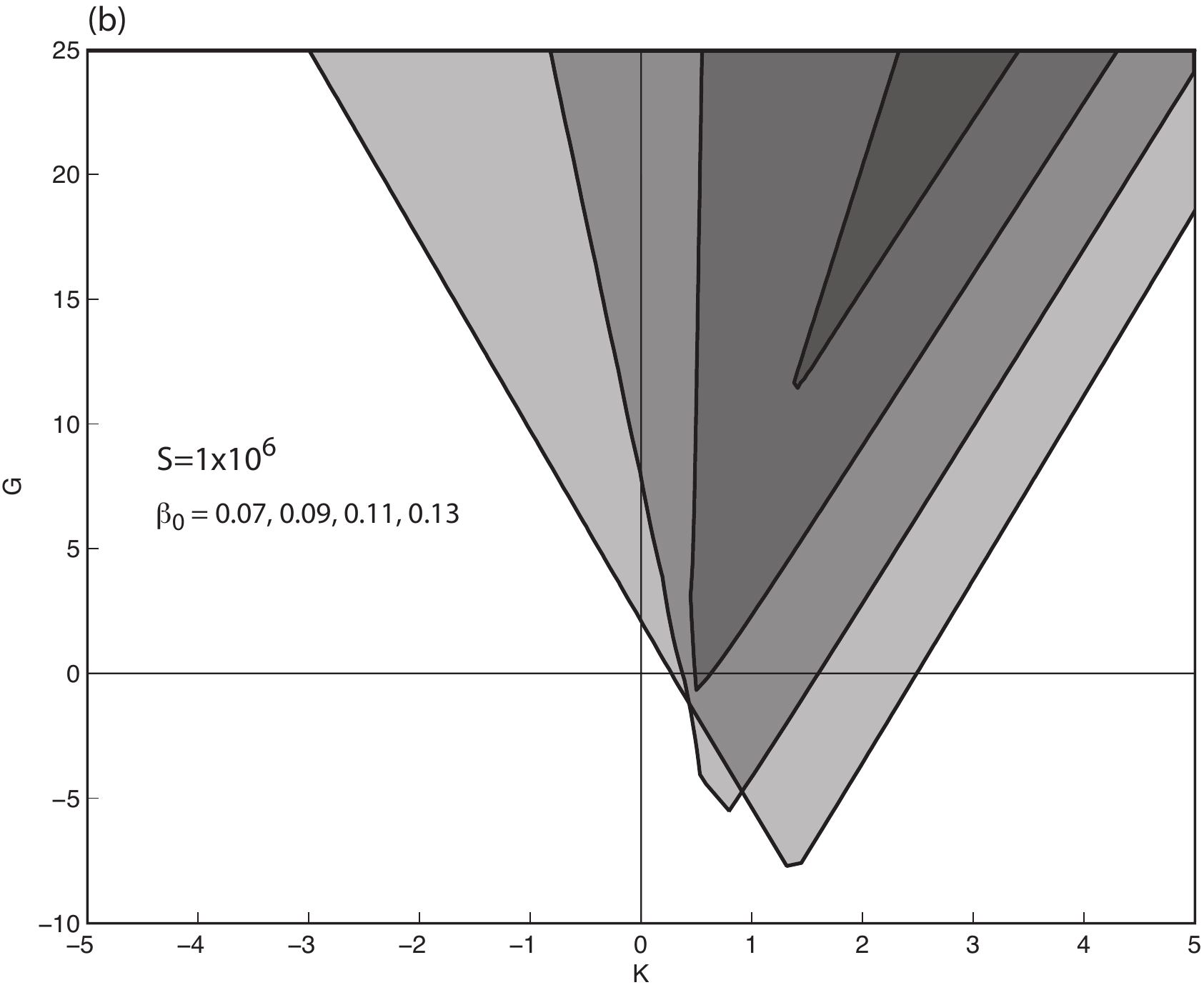}\protect\includegraphics[scale=0.3]{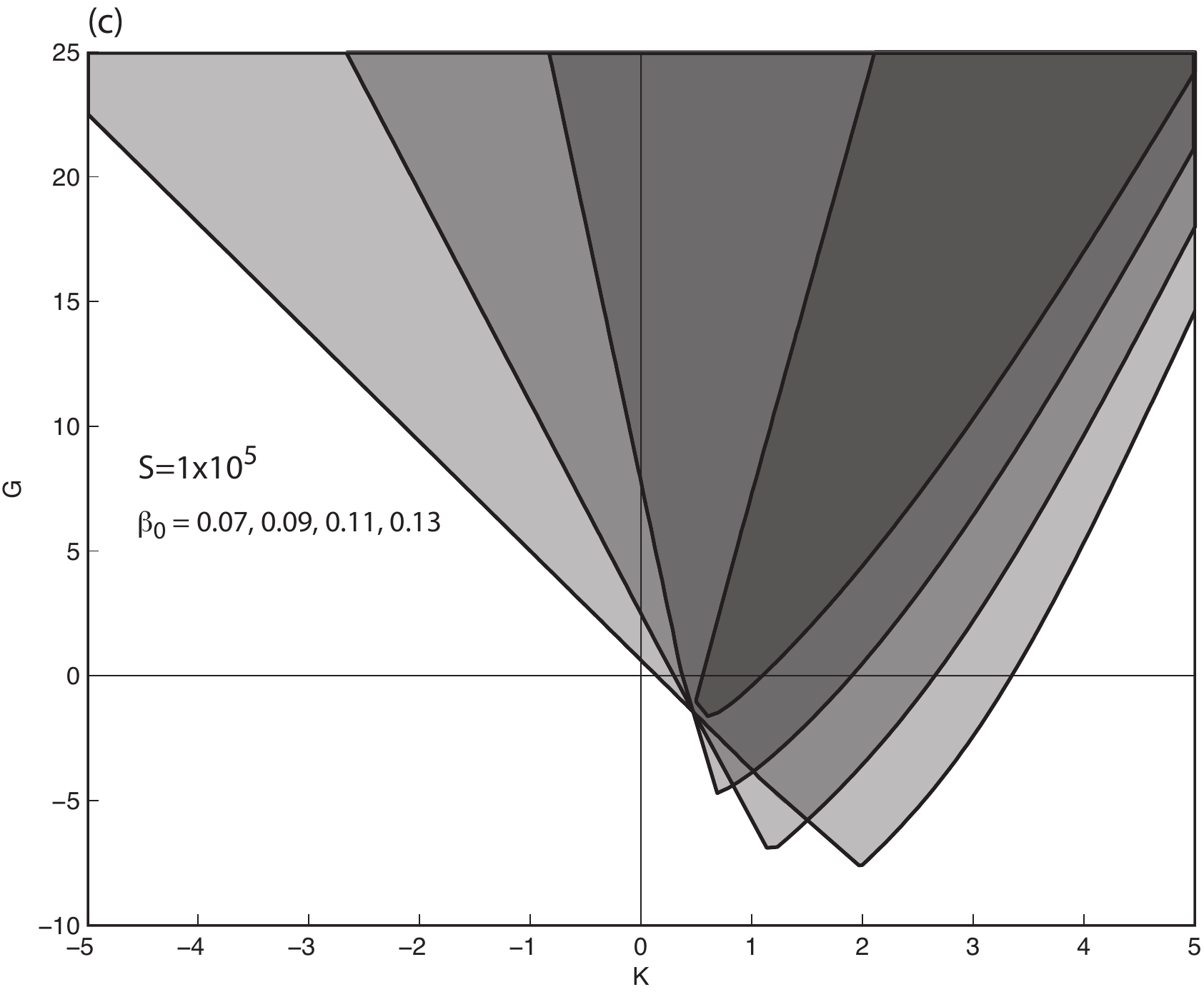}}
\textcolor{black}{Real gain parameter space $(G,K)$ for both the
analytic model (a) and the numerical model (b) and (c), both with
$\Omega=G_{i}=K_{i}=0$. In (a) parameters are as in Fig.~2, with
$\beta_{0}=0.025,\,\,0.075,\,\,0.15$, and $0.225$. The left boundary
is vertical at $\beta_{0}=0.101=\beta_{rp,iw}$. In (b) we have $S=10^{6}$
with $\beta_{0}=0.07,0.09,\,\,0.11=\beta_{rp,iw}$, and $\beta_{0}=0.13$,
and the left boundary is indeed vertical at $\beta_{rp,iw}$. In (c)
we have $S=10^{5}$ }with the same $\beta_{0}$ values \textcolor{black}{as
(b) and the left boundary is also vertical as $\beta_{rp,iw}=0.12$
is crossed. }
\end{figure}
\begin{figure}
\caption{\protect\includegraphics[scale=0.45]{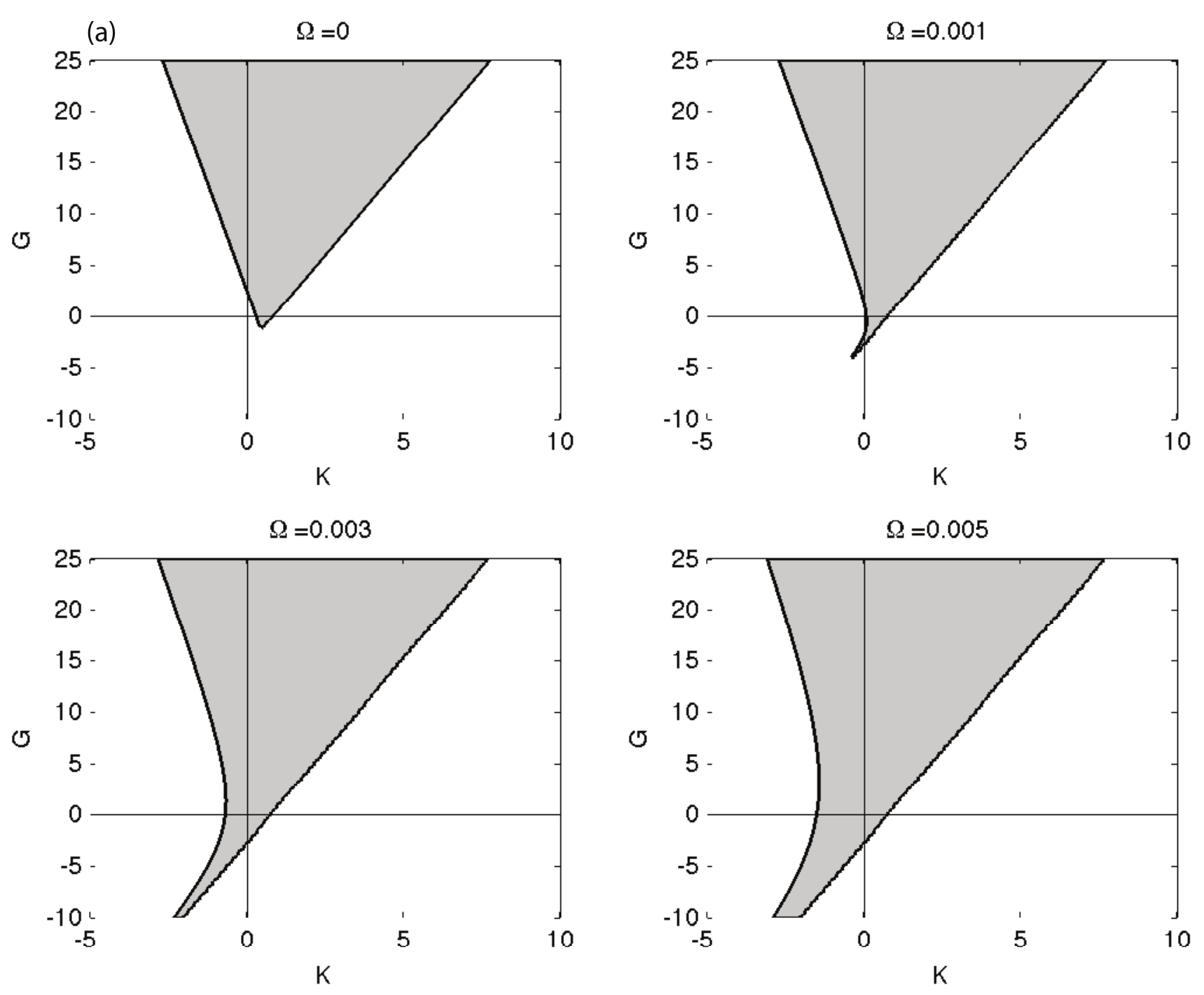}\protect\includegraphics[scale=0.45]{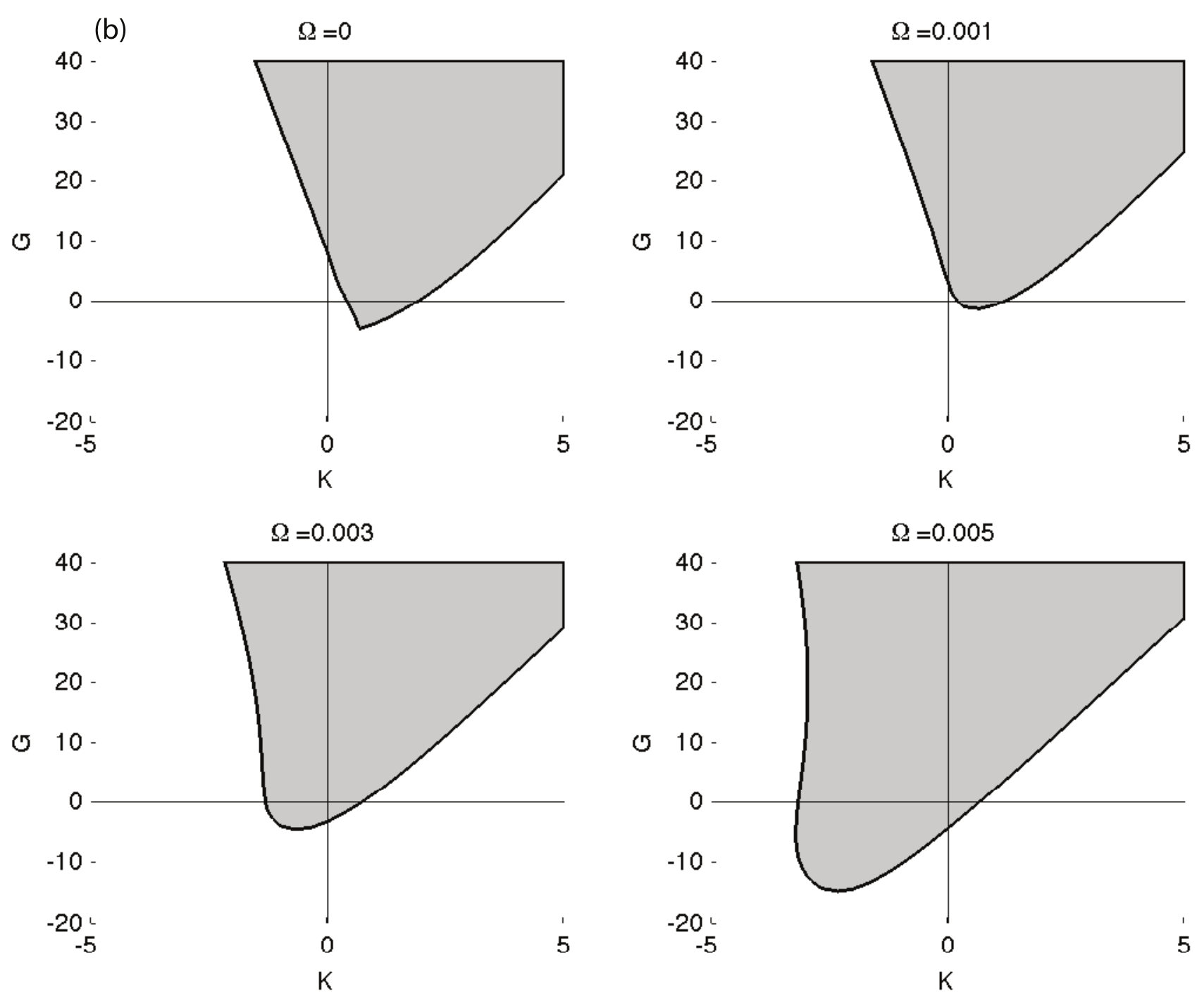}}
\textcolor{black}{Stability diagrams in $(G,K)$ parameter space for
$\beta_{0}<\beta_{rp,iw}$ with plasma rotation $\Omega$ for (a)
the analytic model and (b) the numerical full MHD model. In (a), the
parameters as in Fig.~2 with $\beta_{0}=0.068<\beta_{rp,iw}=0.101$.
In (b), the parameters are as in Fig.~3(b) with $\beta_{0}=0.09\lesssim\beta_{rp,iw}={\color{blue}{\normalcolor 0.12}}$.
The results show that increasing $\Omega$ increases the stable area
for $\beta_{0}<\beta_{rp,iw}$ except for small $\Omega$.}
\end{figure}
\begin{figure}
\caption{\protect\includegraphics[scale=0.3]{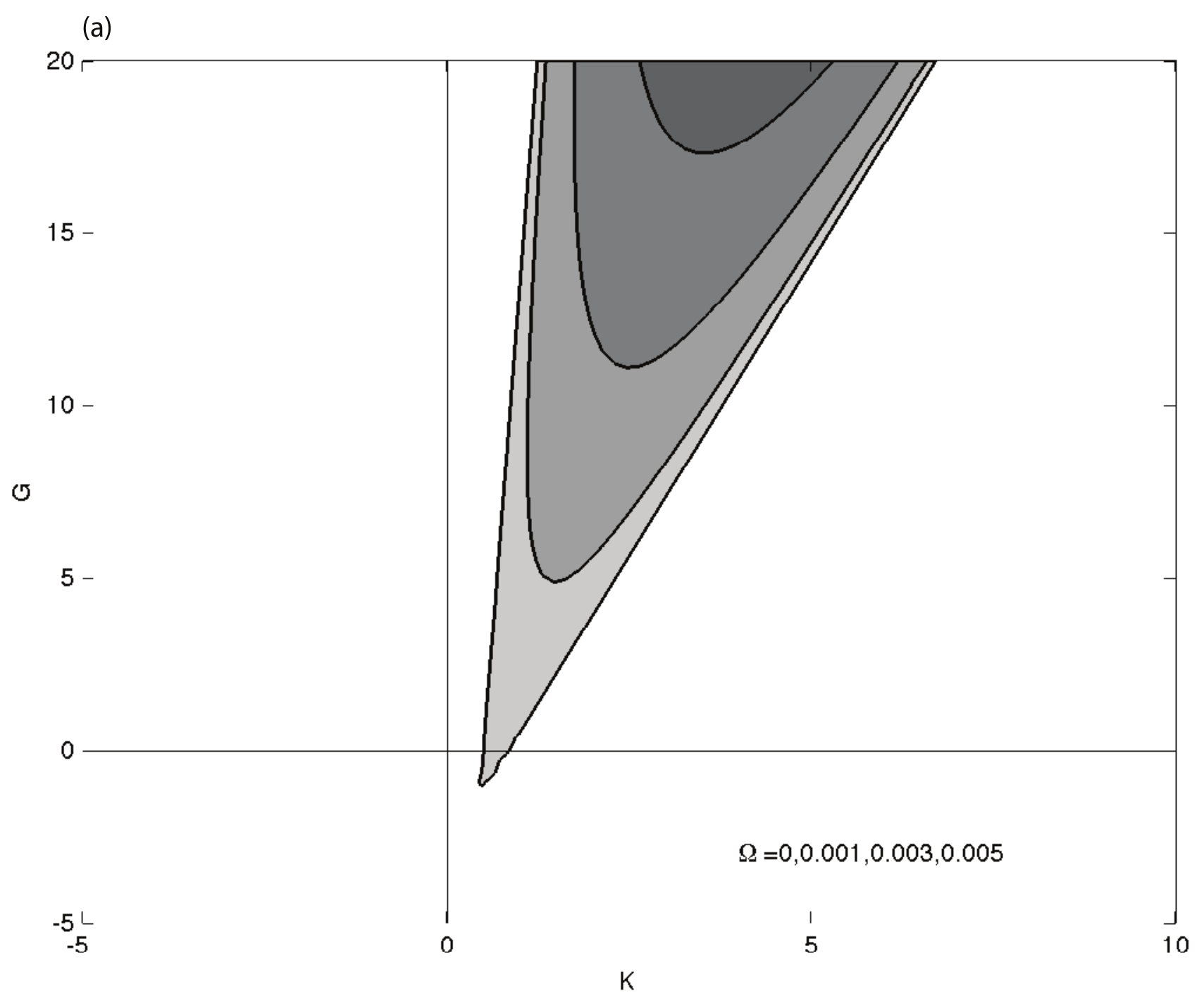}\protect\includegraphics[scale=0.3]{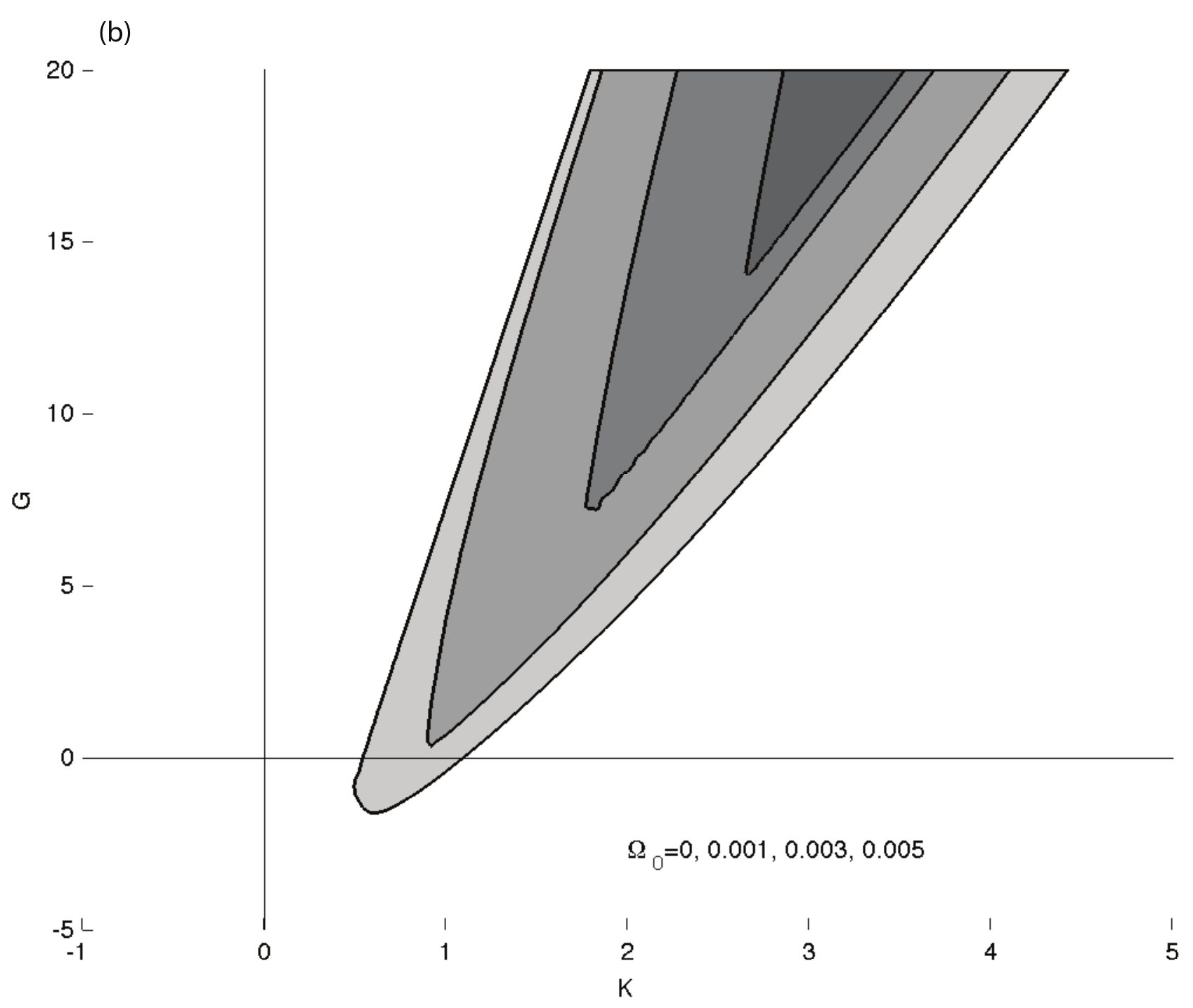}}
\textcolor{black}{Stability diagrams in $(G,K)$ parameter space with
$\beta_{0}>\beta_{rp,iw}$, $G_{i}=K_{i}=0$ and varying rotation
$\Omega$. Shown are (a) simplified model and (b) full MHD model.
In (a) the parameters are as in Fig.~2(a) with $\beta_{0}=0.12$
while in (b) the parameters are as in Fig.~3-4(b) with $\beta_{0}=0.13$.
The plasma Doppler shift frequencies in (a) and (b) are $\Omega=0,\,\,0.001,\,\,0.003,\,0.005$.
These results show that for $\beta_{0}>\beta_{rp,iw}$ the stable
region shrinks as $|\Omega|$ increases.}
\end{figure}
\textcolor{red}{}
\begin{figure}
\caption{\protect\includegraphics[scale=0.45]{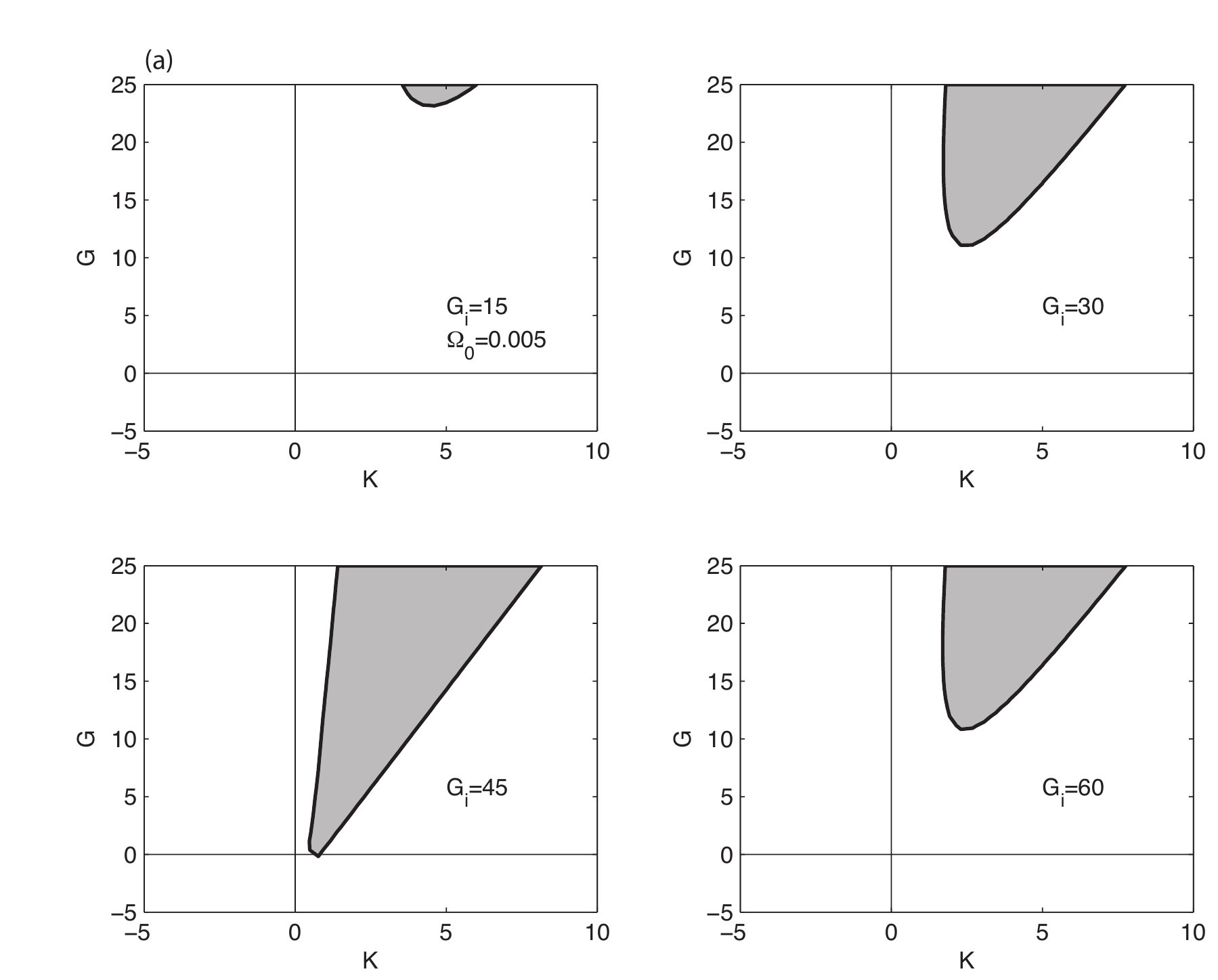}\protect\includegraphics[scale=0.45]{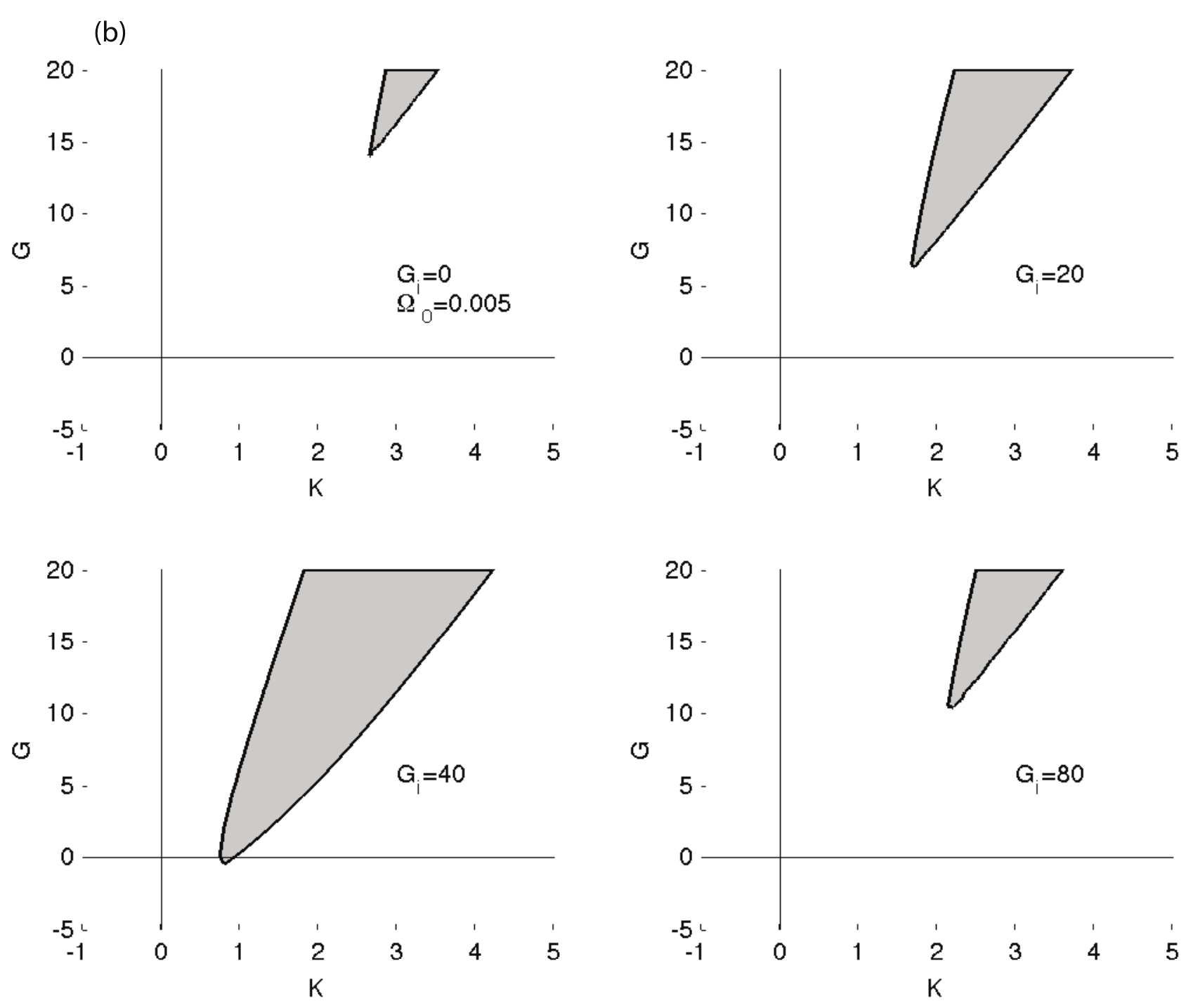}}
\textcolor{black}{Stability diagrams for $\beta_{0}>\beta_{rp,iw}$
and $\Omega\neq0$ for (a) simplified model and (b) full MHD model,
both for $\Omega=0.005$. In (a) the parameters are as in Fig.~2
except for the wall time, which is made equal to the numerical case,}\textcolor{blue}{{}
}having \textcolor{black}{$\tau_{w}=2\times10^{4}$, with $\beta_{0}=0.12>{\color{magenta}{\color{black}\beta}_{{\color{black}{\color{black}}rp,iw}}}$,
and }\textcolor{blue}{${\normalcolor {\normalcolor }G_{i}=15,30,45,60}$.
}\textcolor{black}{In (b) we have $\beta_{0}=0.13>{\color{magenta}{\color{black}\beta}_{{\color{black}{\color{black}}rp,iw}}}$,
with $G_{i}=0,\,\,20,\,\,40,\,\,80$. The results show that there
is an optimal value of $G_{i}$; for this value the effective wall
rotation rate $\Omega_{w}$ is equal to $\Omega$ and the stable region
is maximized. }
\end{figure}
\textcolor{red}{}
\begin{figure}
\caption{\protect\includegraphics[scale=0.45]{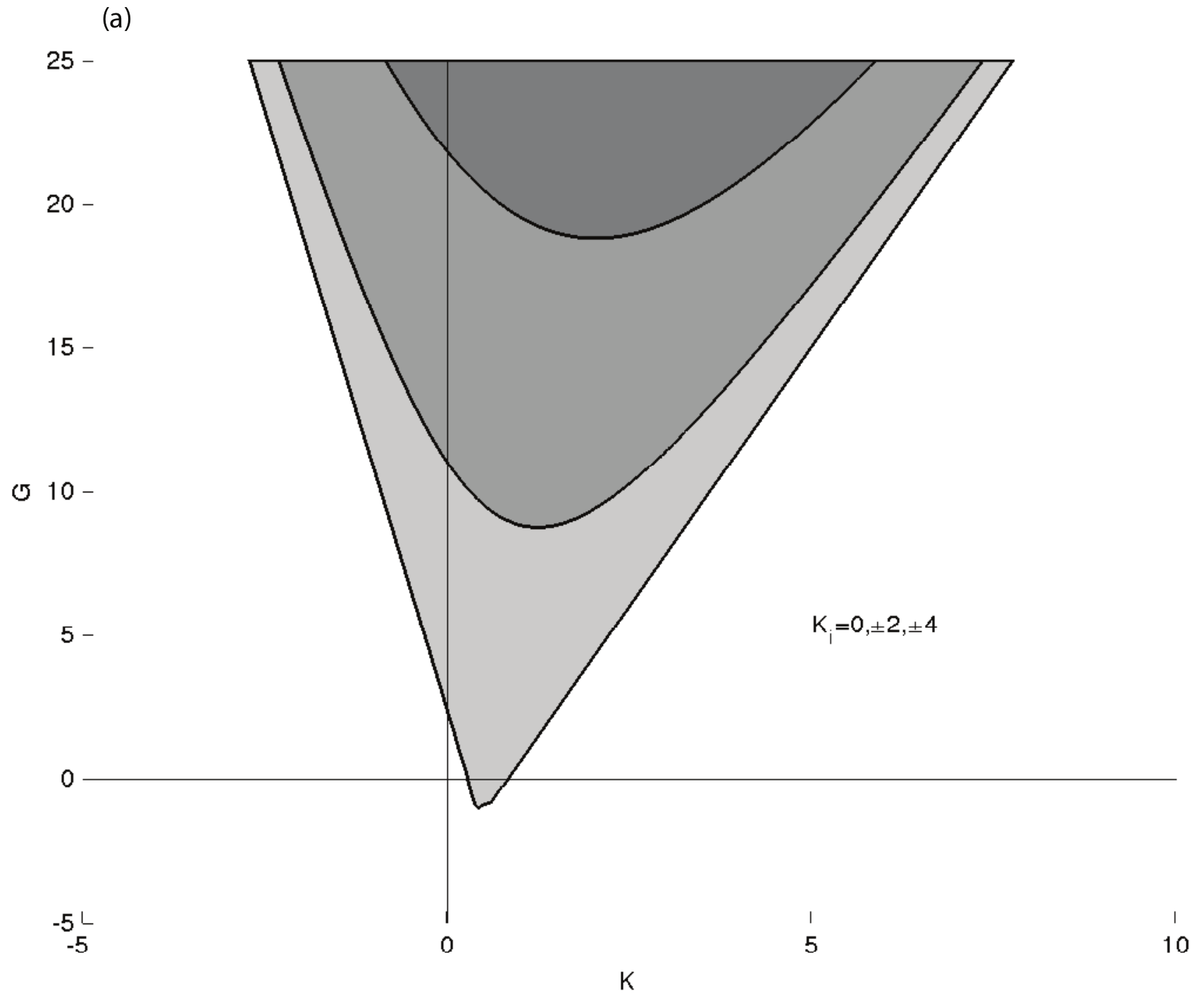}\protect\includegraphics[scale=0.45]{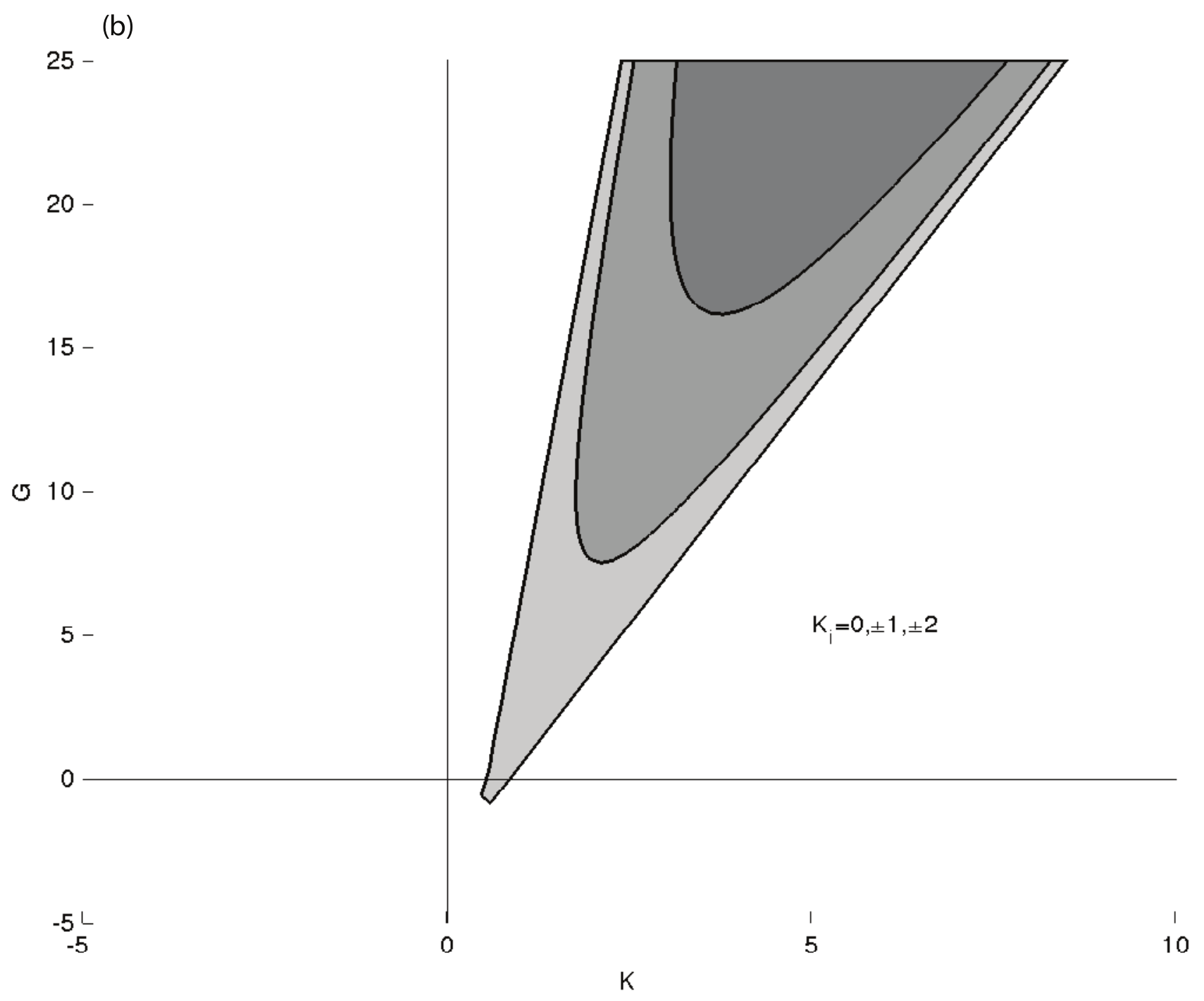}}

\textcolor{black}{Stability diagrams for the simplified model for
(a) $\beta_{0}=0.068<\beta_{rp,iw}$ with $K_{i}=0,\,\,\pm2,\,\,\pm4$
and (b) $\beta_{0}=0.15>\beta_{rp,iw}$ with $K_{i}=0,\,\,\pm1,\,\,\pm2$.
In both regimes of $\beta_{0}$, $K_{i}$ decreases the size of the
stable region.}
\end{figure}
\begin{figure}
\caption{\protect\includegraphics[scale=0.45]{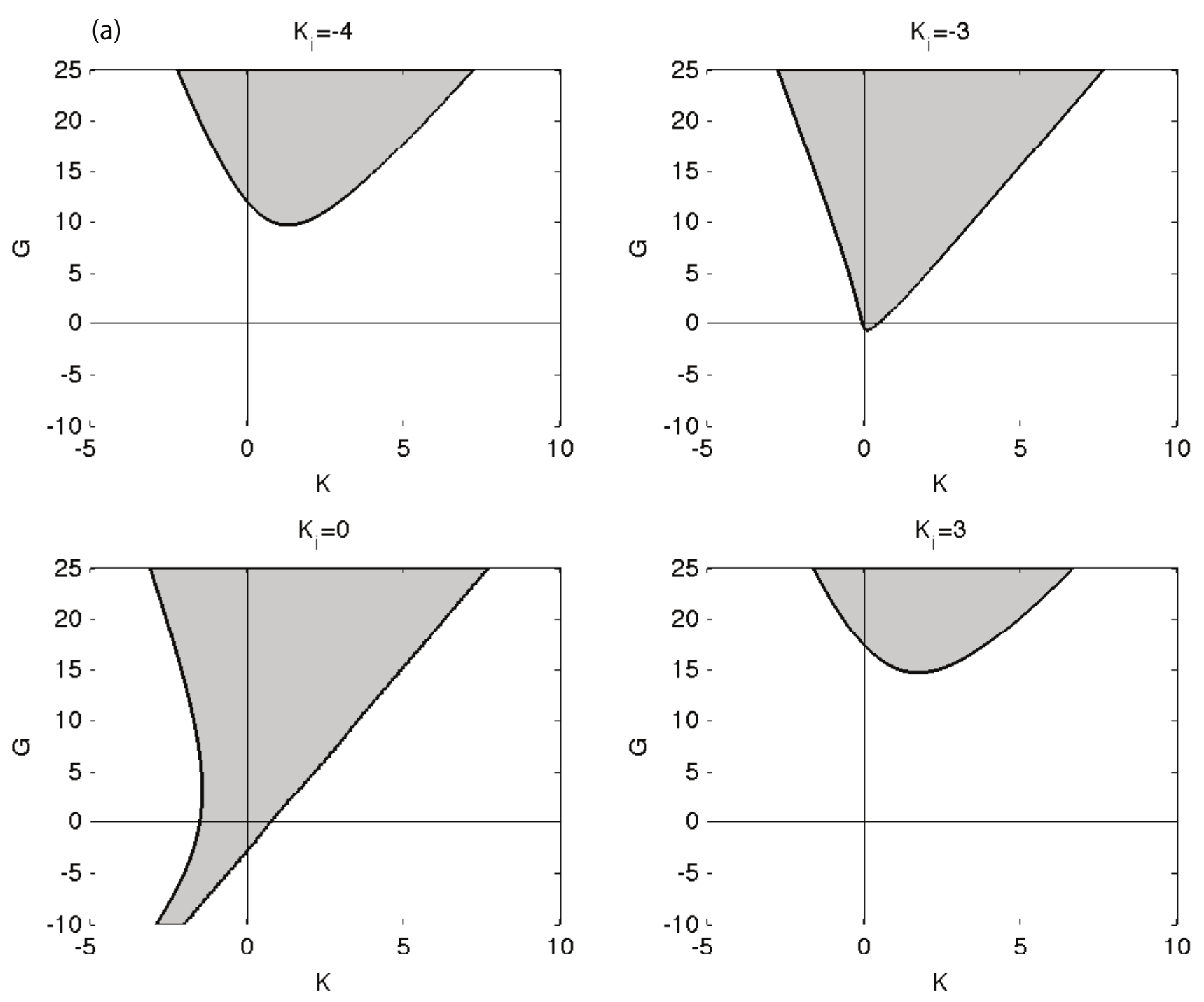}\protect\includegraphics[scale=0.45]{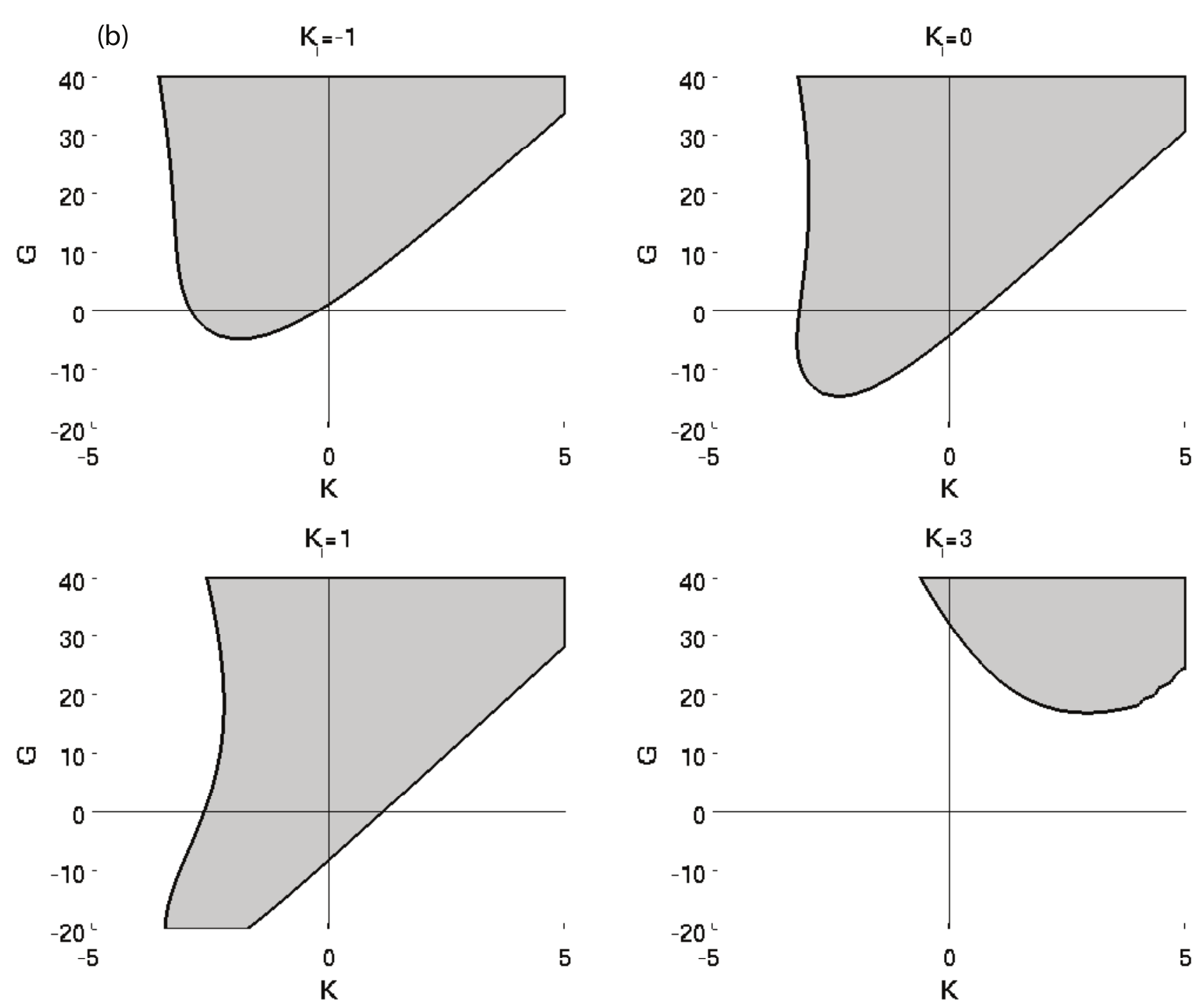}}
\textcolor{black}{Stability diagrams with $G=0$ and varying $K_{i}$
for $\beta_{0}<\beta_{rp,iw}$ and $\Omega=0.005$, with (a) the simplified
model with $\beta_{0}=0.068$ and (b) the full MHD model with $\beta_{0}=0.09$.
These results show that the optimal value of $|K_{i}|$ is small and
larger values destabilize in the $\beta_{0}<\beta_{rp,iw}$ regime. }
\end{figure}
\begin{figure}
\caption{\protect\includegraphics[scale=0.45]{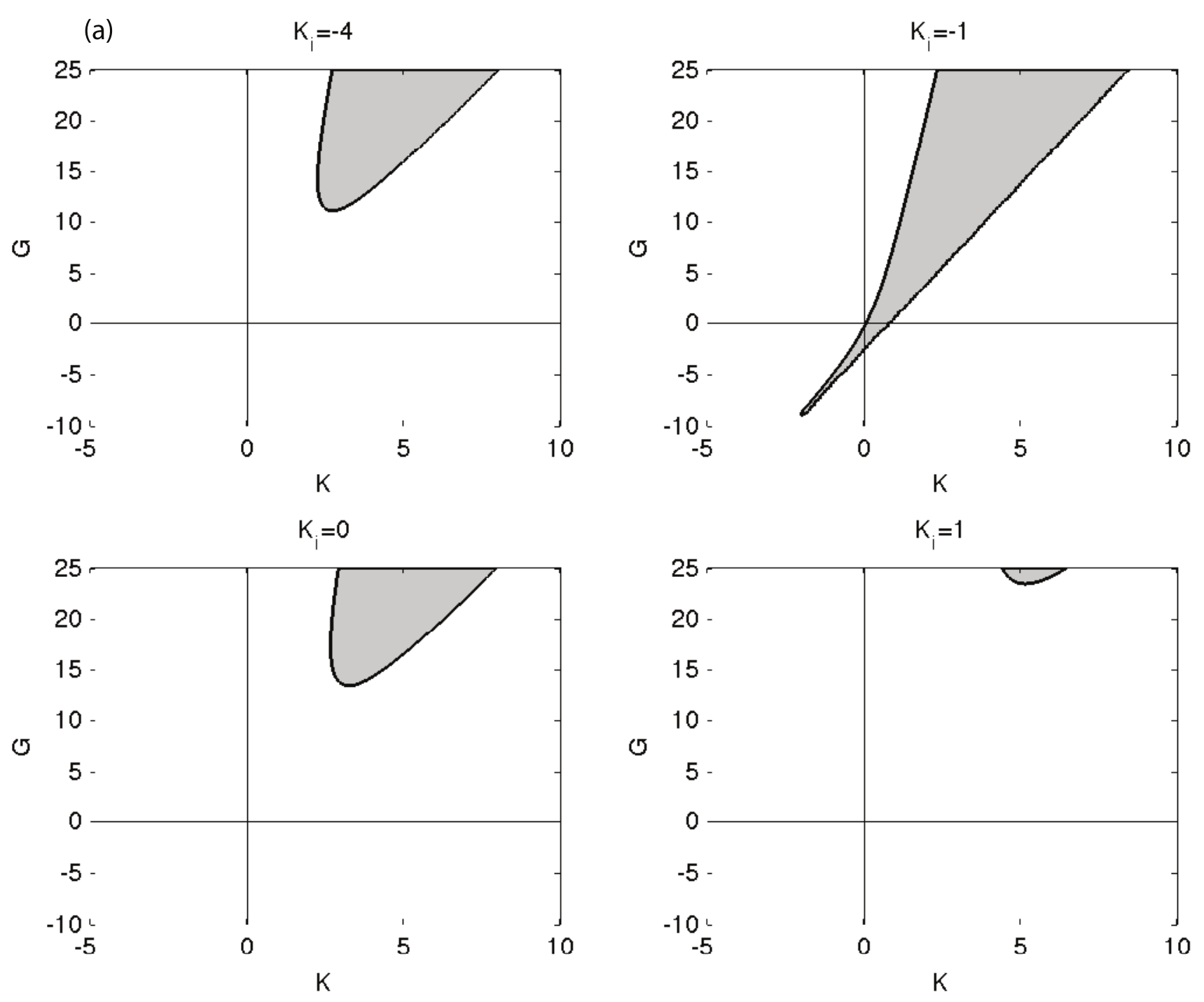}\protect\includegraphics[scale=0.45]{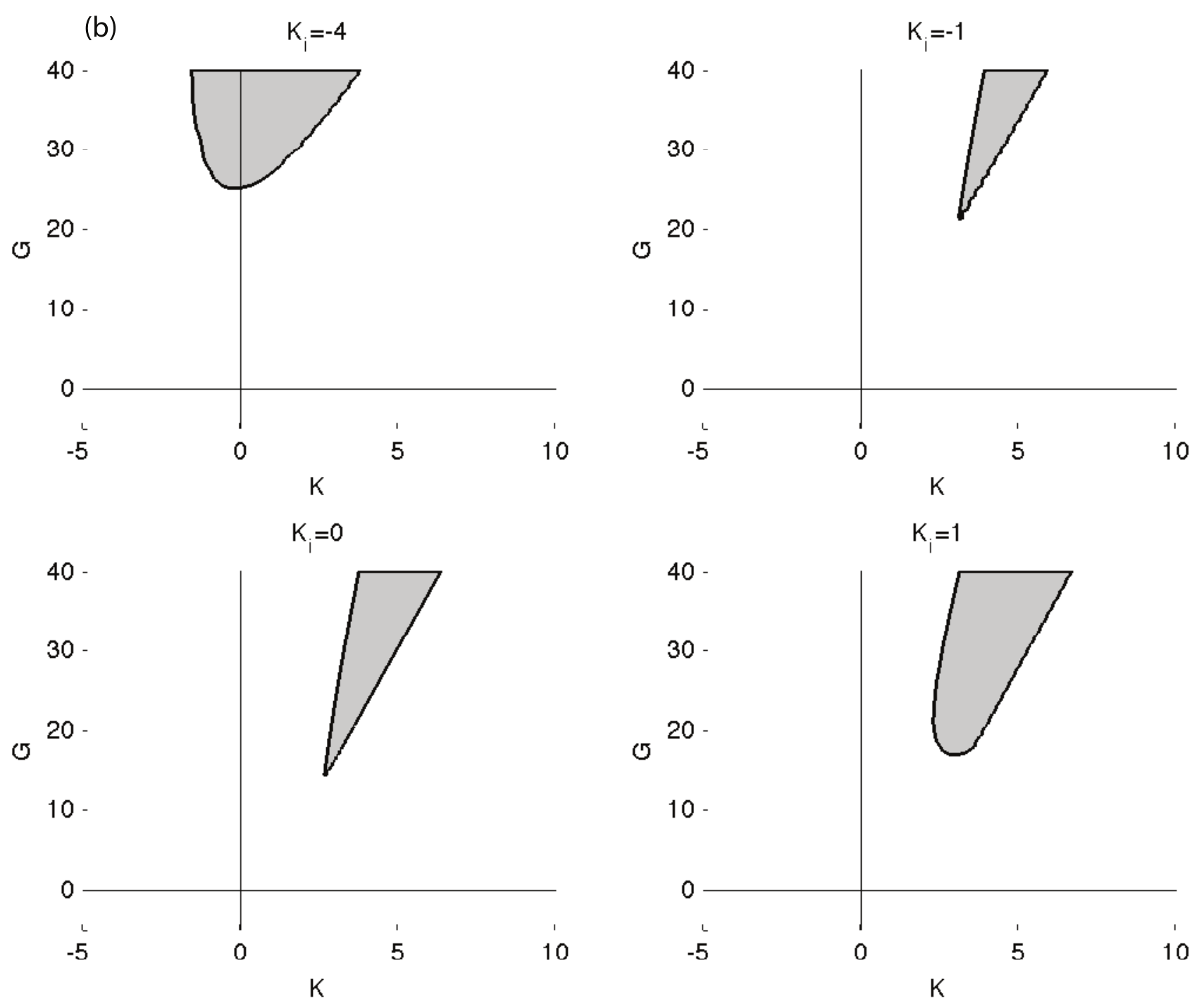}}
\textcolor{black}{Stability diagram with $\beta_{0}>\beta_{rp,iw}$,
$\Omega=0.005,\,\, G_{i}=0$ and $K_{i}=-4,-1,0$ and $1$ for (a)
the simplified model with $\beta_{0}=0.15$ and (b) full MHD model
with $\beta_{0}=0.13$ as in Fig.~5(b). }In (a) the optimal value
of $K_{i}$ is -1. In (b) the stability regions are more complex,
but optimal for $K_{i}$ for small $K_{i}$.
\end{figure}
\begin{figure}
\caption{\protect\includegraphics[scale=0.25]{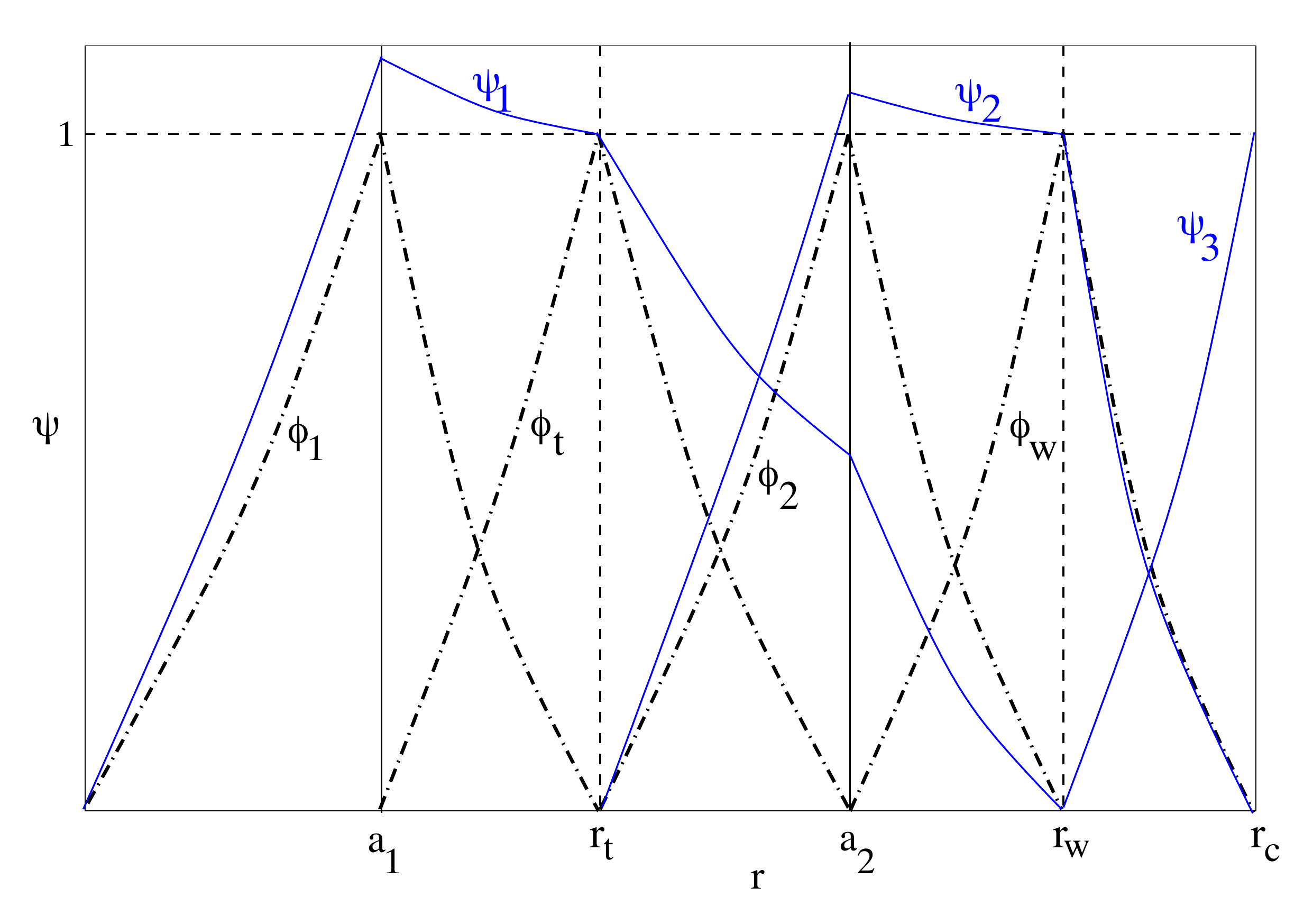}}
Sketch of basis functions $\phi_{1},\,\phi_{t},\,\phi_{2},\,\phi_{w}$
used to derive functions $\psi_{1},\,\psi_{2}$, and $\psi_{3}$ of
the Appendix, also shown.
\end{figure}

\end{document}